\documentclass[floatfix,a4paper,tightenlines,amsmath,amssymb,nofootinbib,showpacs,notitlepage,showkeys,prd,twocolumn,10pt]{revtex4-1}

\usepackage[utf8]{inputenc}
\usepackage{graphicx}
\usepackage{hyperref}

\newcommand{\fref}[1]{Fig. \ref{#1}}
\newcommand{\nnnl}{\nonumber\\}	

\allowdisplaybreaks 

\begin{document}

\title{Correlation functions of three-dimensional Yang-Mills theory from Dyson-Schwinger equations}

\author{Markus Q.~Huber}
\affiliation{Institute of Physics, University of Graz, NAWI Graz, Universit\"atsplatz 5, 8010 Graz, Austria}
\email{markus.huber@uni-graz.at}
\date{\today}

\begin{abstract}
The two- and three-point functions and the four-gluon vertex of three-dimensional Yang-Mills theory are calculated from their Dyson-Schwinger equations and the 3PI effective action. Within a self-contained truncation various effects of truncating Dyson-Schwinger equations are studied. Estimates for the errors induced by truncations are derived from comparisons between results from different equations, comparisons with lattice results, and varying higher Green functions. The results indicate that the two-loop diagrams are important in the gluon propagator, where they are explicitly calculated, but not for the vertices. Furthermore, the influence of the four-gluon vertex on lower Green functions is found to be small.
\end{abstract}



\maketitle

\section{Introduction}

At low momentum transfer quantum chromodynamics (QCD) is a strongly coupled theory. To describe its rich nonperturbative phenomenology, e.g., bound states with a dynamically generated mass and with constituents which cannot be observed as free particles, corresponding methods are required.

One nonperturbative approach in the continuum is functional equations like Dyson-Schwinger equations, see, e.g., Refs.~\cite{Roberts:1994dr,Alkofer:2000wg,Alkofer:2008nt}, the functional renormalization group, see, e.g., Refs.~\cite{Berges:2000ew,Pawlowski:2005xe,Gies:2006wv} or equations of motion from n-PI effective actions \cite{Cornwall:1974vz,Berges:2004pu}.
Applications of functional equations in QCD are manifold and include the description of hadrons, see, e.g., Refs.~\cite{Alkofer:2000wg,Bashir:2012fs,Eichmann:2013afa}, investigations of QCD at high density, e.g., \cite{Muller:2013pya,Muller:2013tya} and studies of the chiral and deconfinement transitions, e.g., \cite{Braun:2007bx,Braun:2009gm,Fister:2011uw,Fischer:2011mz,Fischer:2012vc,Fister:2013bh,Haas:2013qwp,Fischer:2013eca,Fischer:2014ata}.
In such calculations varying quark masses down to the chiral limit or introducing a chemical potential poses no principal obstacle. The drawback is that functional equations consist of infinitely large sets of equations which must be truncated for most applications. Typically, it is difficult to assess the effect a certain truncation has on the results. In this work, several tests are suggested and applied.

Functional equations are coupled, (non)linear integral or integro-differential equations formulated in terms of the correlation functions of a theory. To calculate a specific set of correlation functions, the nonincluded correlation functions must be specified by a model. This includes the possibility of setting it to zero and all diagrams that contain it can be dropped. In general, a truncation is thus specified by (I) the set of correlation functions calculated dynamically, by (II) the set of dropped correlation functions and by (III) models for all others.

Unfortunately, there is no general hierarchy of diagrams valid for all momenta on which to base the construction of truncations. For example, perturbation theory of course provides an ordering scheme in terms of the coupling constant $g$ valid for high momenta, but in the nonperturbative regime this is no longer applicable. Indeed, examples are known in which in the deep IR two-loop diagrams become more important than one-loop diagrams \cite{Huber:2009wh}. In particular, for the midmomentum regime it is difficult to assess the importance of single diagrams.

For a given truncation the challenge is thus to assess the induced error. A straightforward possibility is a comparison with results from another method like lattice calculations if they are available. However, typically some model dependence remains, and comparisons have to be taken with a grain of salt. This also applies to comparisons between different truncations.

Nevertheless, by enlarging truncations one can still learn something about the importance of different sectors of the theory. To quote only one example, it was found that in the Yang-Mills sector of QCD nonperturbatively generated dressings for vertices \cite{Blum:2014gna,Eichmann:2014xya,Cyrol:2014kca} are less important than in the matter sector where perturbatively absent dressings are produced by the dynamical breaking of chiral symmetry \cite{Windisch:2014th,Hopfer:2014th,Williams:2014iea,Mitter:2014wpa,Williams:2015cvx}.

While the explicit meaning of \textit{truncation} is to neglect certain diagrams or model some correlation functions to decouple a closed set of equations from the infinite system, it has to be noted that, strictly speaking, also other modifications of the equations have to be taken into account. Some of them seem purely technical, but they still can have an impact on the solution. A prime example is certainly spurious divergences in the gluon propagator Dyson-Schwinger equation (DSE) and how they are subtracted \cite{Huber:2014tva}. Another example is renormalization group (RG) improvement terms that are added to obtain the correct resummed perturbative behavior \cite{vonSmekal:1997vx,Huber:2012kd}.

\begin{figure}[tb]
 \begin{center}
  \includegraphics[width=0.38\textwidth]{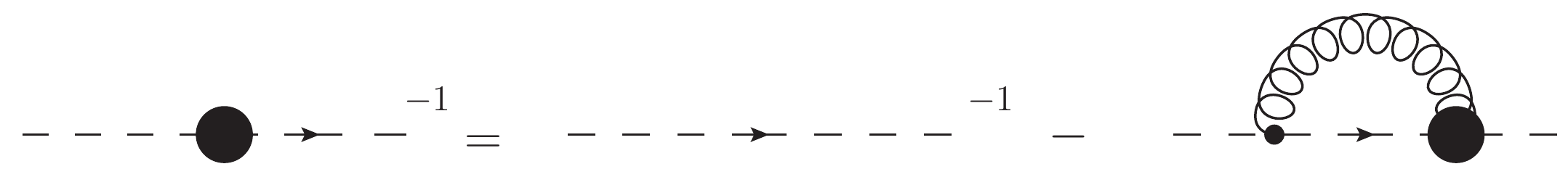}
  \caption{\label{fig:gh-DSE}The ghost propagator Dyson-Schwinger equation. Here and in all other figures, internal propagators are dressed, and thick blobs denote dressed vertices, wiggly lines gluons, and dashed ones ghosts.}
 \end{center}
\end{figure}

\begin{figure}[tb]
 \begin{center}
  \includegraphics[width=0.45\textwidth]{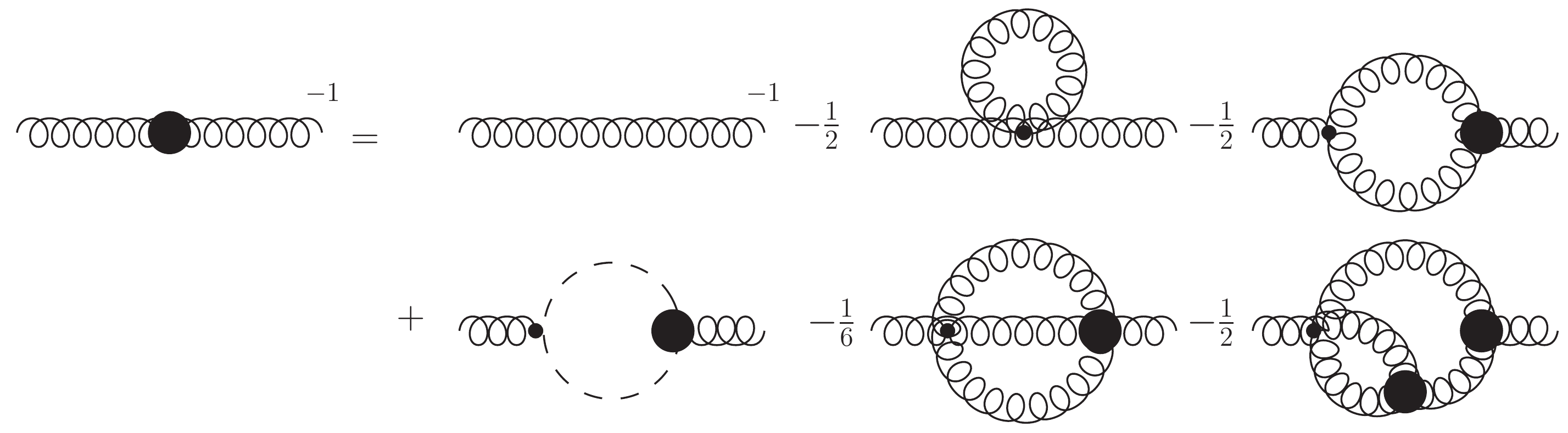}
  \caption{\label{fig:gl-DSE}The gluon propagator Dyson-Schwinger equations. The loop diagrams in the gluon propagator DSE are called the tadpole, gluon loop, ghost loop, sunset, and squint.}
 \end{center}
\end{figure}

In this work, the DSEs and equations of motion from the 3PI effective action of three-dimensional Yang-Mills theory are investigated. Since this theory is finite, no renormalization is necessary and the handling of spurious divergences is easier compared to the four-dimensional theory. A truncation that comprises all five primitively divergent correlation functions is introduced in Sec.~\ref{sec:DSEs}. Sec.~\ref{sec:YM3d} contains a short review of three-dimensional Yang-Mills theory to which this truncation is applied and then tested in several ways. In Secs.~\ref{sec:ghg} and \ref{sec:4g_effects}, the influences of the choice of equations and of the highest included correlation function, the four-gluon vertex, are investigated. Sec.~\ref{sec:results} contains a comparison with lattice results. Sec.~\ref{sec:summary} contains some concluding remarks. The appendices contain additional information on technical details and the handling of spurious divergences.

\section{Equations of Yang-Mills theory and their truncations}
\label{sec:DSEs}

Dyson-Schwinger equations (DSEs) can be derived from the invariance of the path integral under translations of the fields, see, e.g., Refs.~\cite{Roberts:1994dr,Alkofer:2000wg,Alkofer:2008nt}. In this work, the DSEs for the propagators, the three-point functions, and the four-gluon vertex are investigated. These quantities are parametrized as follows. The gluon propagator, being completely transverse in the Landau gauge, is denoted by
\begin{align}
D_{\mu\nu}^{ab}(p)=\frac{\delta^{ab}}{p^2}Z(p^2)P_{\mu\nu}(p)
\end{align}
where $P_{\mu\nu}(p)=g_{\mu\nu}-p_\mu p_\nu/p^2$ is the transverse projector. The gluon two-point function, defined as
\begin{align}
\Gamma_{\mu\nu}^{ab}(p)&=\delta^{ab}\Gamma(p^2)P_{\mu\nu}(p) p^2,
\end{align}
is the inverse propagator, so $\Gamma(p^2)=1/Z(p^2)$. The ghost propagator is given by
\begin{align}
 D^{ab}(p)&=-\delta^{ab}G(p^2)\frac1{p^2}.
\end{align}

\begin{figure}[tb]
 \begin{center}
  \hskip2cm\includegraphics[width=0.45\textwidth]{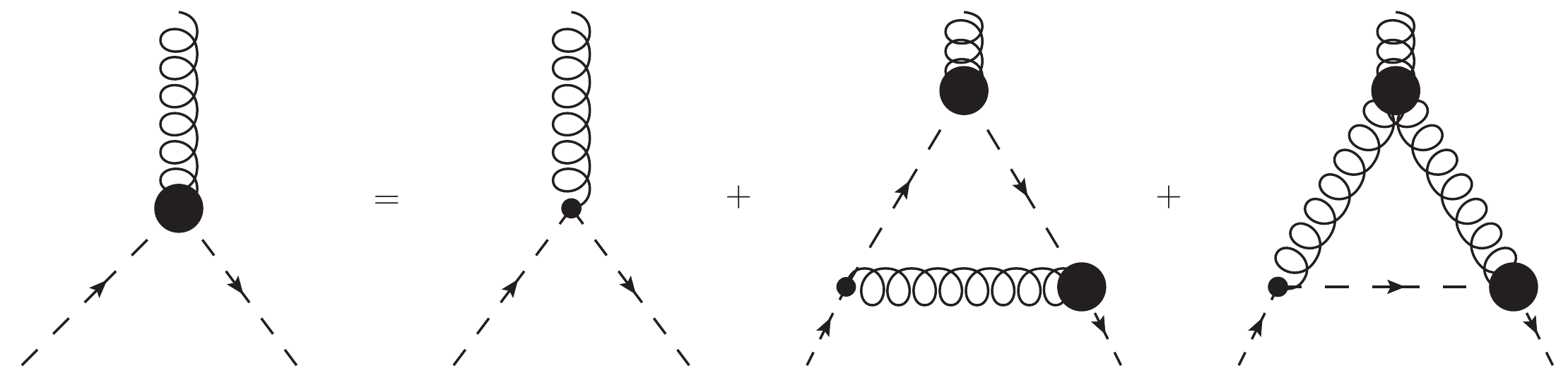}\\
  \vskip2mm
  \hskip2cm\includegraphics[width=0.45\textwidth]{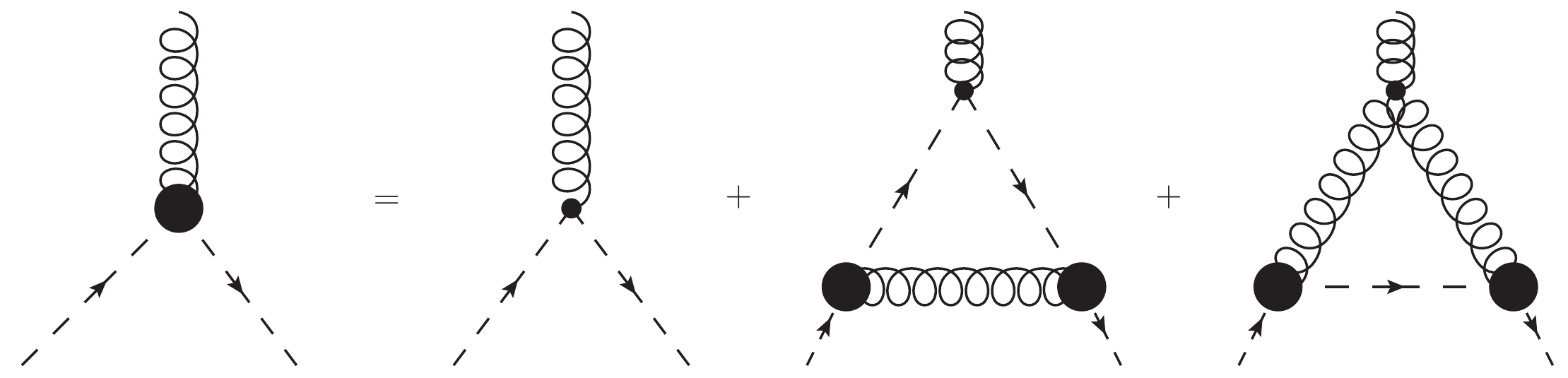}
  \caption{\label{fig:ghg-DSE}Top: Truncated $c$-DSE of the ghost-gluon vertex. Bottom: Truncated $A$-DSE of the ghost-gluon vertex. The full DSEs can be found, for example, in Ref.~\cite{Alkofer:2008nt}. The loop diagrams are called Abelian and non-Abelian triangles.}
 \end{center}
\end{figure}

The ghost-gluon and three-gluon vertices are denoted by
\begin{align}
&\Gamma_\mu^{abc}(k;p,q)=\nnnl
&i\,g f^{abc}\left(A(k^2;p^2,q^2)p_\mu+B(k^2;p^2,q^2)k_\mu\right),\\
&\Gamma_{\mu\nu\rho}^{abc}(p,q,r)=\nnnl
&\,i\,g\,f^{abc}D^{A^3}(p^2,q^2,r^2)\left( (r-q)_\mu g_{\nu\rho} + \text{perm.} \right),
\end{align}
respectively. For the ghost-gluon vertex only the dressing function $A(k^2; p^2, q^2)$ will be calculated as $B(k^2;p^2,q^2)$ drops out in the Landau gauge. The expression for the three-gluon vertex was already reduced to the tree-level tensor and its dressing. The reason behind this is that other dressings are known to be small as discussed below.

Finally, the four-gluon vertex, also reduced to the tree-level tensor, is parametrized as
\begin{align}
& \Gamma^{abcd}_{\mu\nu\rho\sigma}(p,q,r,s)=\nnnl
&\quad g^2 D^{A^4}(p,q,r,s)\Big( (g_{\nu\rho}g_{\mu\sigma}- g_{\mu\rho}g_{\nu\sigma} )f^{abi}f^{cdi} \nnnl
&\quad+(g_{\mu\sigma}g_{\nu\rho}-g_{\mu\nu}g_{\rho\sigma}) f^{aci}f^{bdi}\nnnl
&\quad +(g_{\mu\rho}g_{\nu\sigma}-g_{\mu\nu}g_{\rho\sigma})f^{adi}f^{bci} \Big). 
\end{align}

\begin{figure}[tb]
 \begin{center}
  \includegraphics[width=0.45\textwidth]{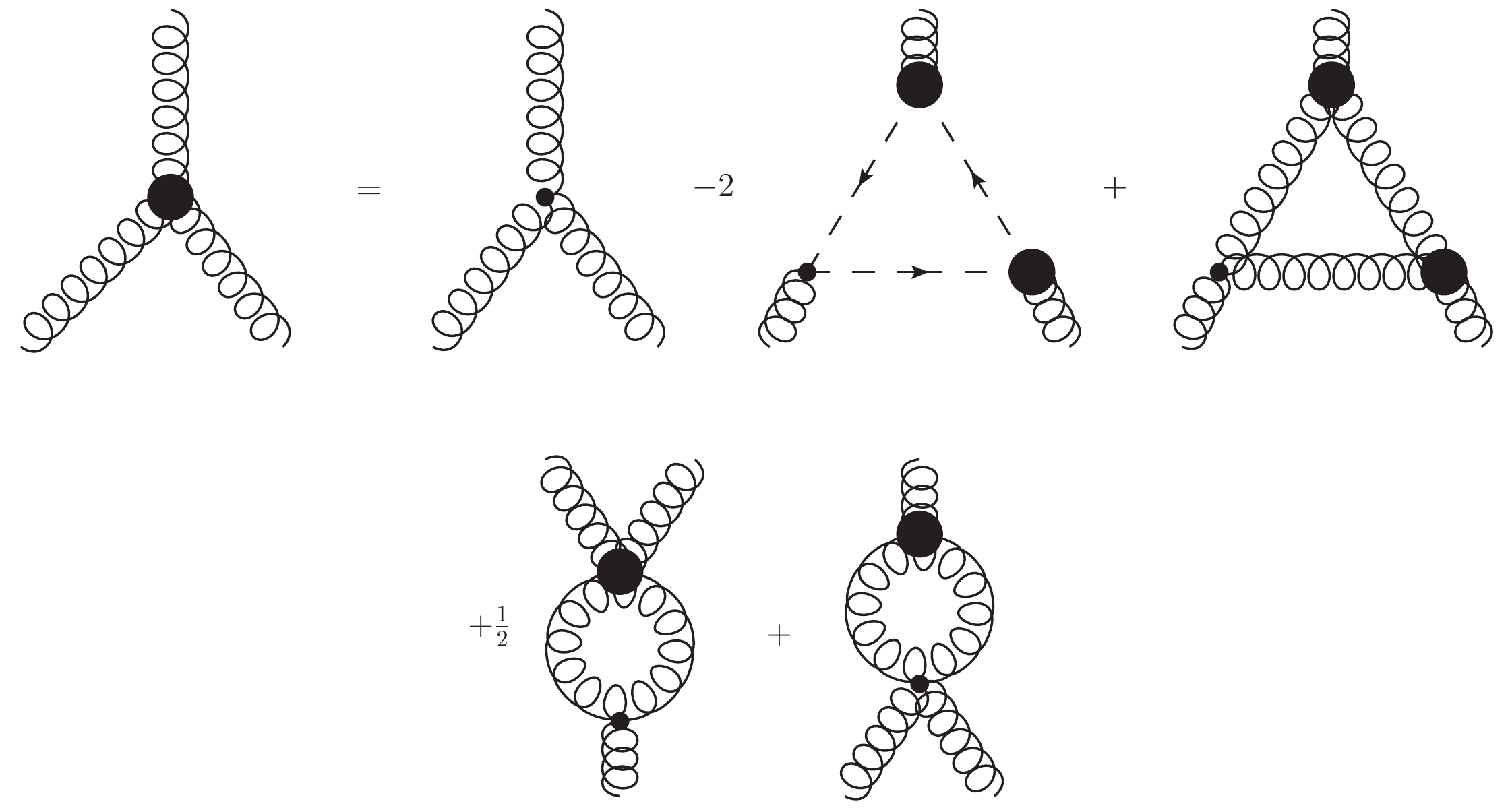}
  \caption{\label{fig:3g-DSE}Truncated DSE of the three-gluon vertex. The full one can be found, e.g., in Ref.~\cite{Huber:2012zj}. The loop diagrams are called the ghost triangle, gluon triangle, static swordfish, and dynamic swordfish. The dynamic swordfish is required only once (with a modified symmetry factor) because of the symmetrization applied to the results.}
 \end{center}
\end{figure}

\begin{figure}[tb]
 \begin{center}
  \includegraphics[width=0.45\textwidth]{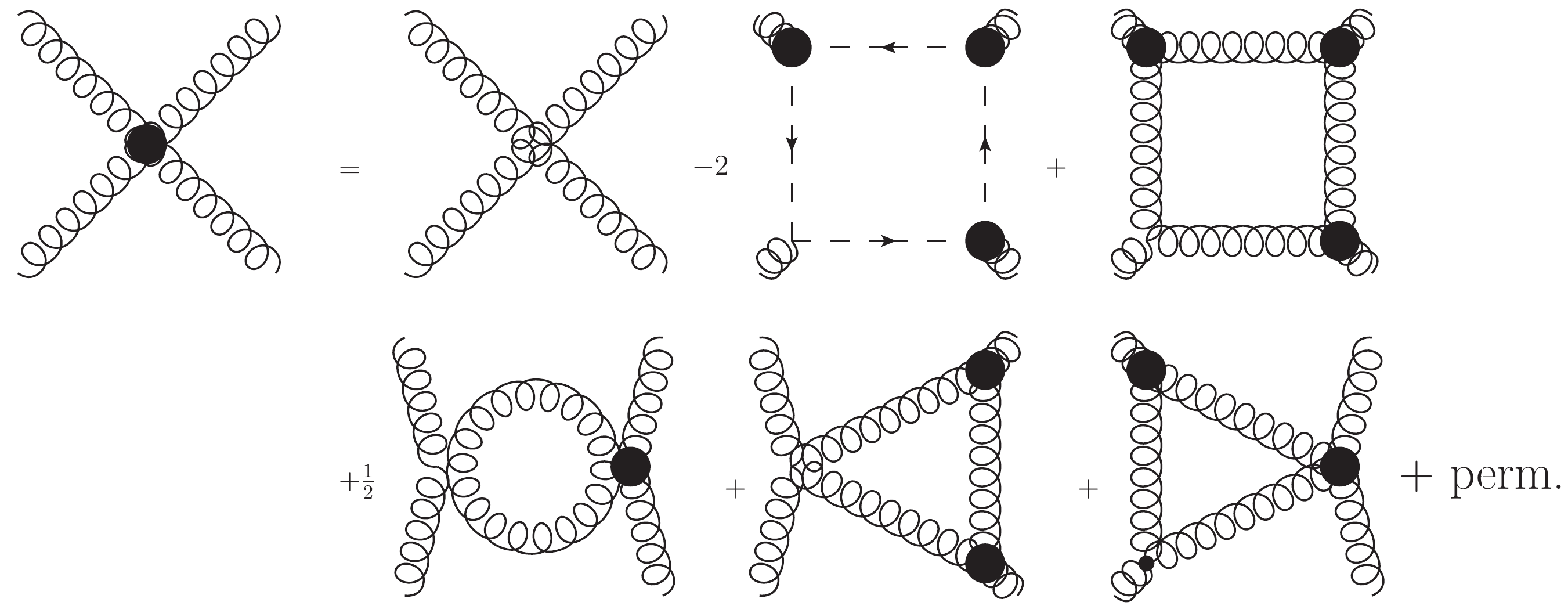}
  \caption{\label{fig:4g-DSE}Truncated DSE of the four-gluon vertex. The loop diagrams are called the ghost box, gluon box, swordfish, static triangle, and dynamic triangle.}
 \end{center}
\end{figure}

The Dyson-Schwinger equations for the propagators are depicted in Figs.~\ref{fig:gh-DSE} and \ref{fig:gl-DSE}. For the ghost-gluon vertex two equations exist: one in which the bare vertex that appears in each DSE is connected to the ghost and one in which it is connected to the gluon legs.\footnote{Of course there exists a variant where the antighost leg is connected to the bare vertex. However, due to the ghost/antighost symmetry of the Landau gauge \cite{Alkofer:2000wg,Lerche:2002ep} this equation is basically identical to the one with the ghost at the bare vertex.} For easy reference we will call the former $A$-DSE and the latter $c$-DSE. Fig.~\ref{fig:ghg-DSE} shows the truncated equations for the $A$-DSE and the $c$-DSE. The truncated equations for the three- and four-gluon vertices are depicted in Figs.~\ref{fig:3g-DSE} and \ref{fig:4g-DSE}.

Truncating a DSE violates Bose symmetry because of the bare vertex contained in each diagram of DSEs. This is taken into account by symmetrizing the results of the equations; see Refs.~\cite{Blum:2014gna,Eichmann:2014xya,Cyrol:2014kca} for details. In the three-gluon vertex DSE, shown in \fref{fig:3g-DSE}, this entails that only two of three swordfish diagrams need to be calculated if one prefactor is adapted, while for the four-gluon vertex, depicted in \fref{fig:4g-DSE}, only 5 instead of 15 diagrams need to be calculated.

To obtain the truncated equations shown in the figures, the following truncation prescription is used: All primitively divergent Green functions are kept and nonprimitively divergent Green functions are set to zero.\footnote{The expression \textit{primitively divergent} is used in analogy to four dimensions although all Green functions are finite in three dimensions.} Consequently, the propagator DSEs, which contain only primitively divergent Green functions, are not truncated. In the three- and four-point function DSEs all UV leading diagrams are retained. To complete the prescription, a DSE for the ghost-gluon vertex has to be chosen. Consequences of that choice are discussed in Sec.~\ref{sec:ghg}. Note that the tadpole diagram in the gluon propagator DSE is not dropped. In most variants of subtracting spurious divergences the tadpole diagram is regarded as a pure quadratic divergence. However, with the method employed here, its finite contributions can be calculated. The fact that it should not be discarded is furthermore emphasized by the role it plays for implicit regularization to maintain gauge invariance \cite{Sampaio:2005pc}.

\begin{figure}[tb]
 \begin{center}
  \includegraphics[width=0.45\textwidth]{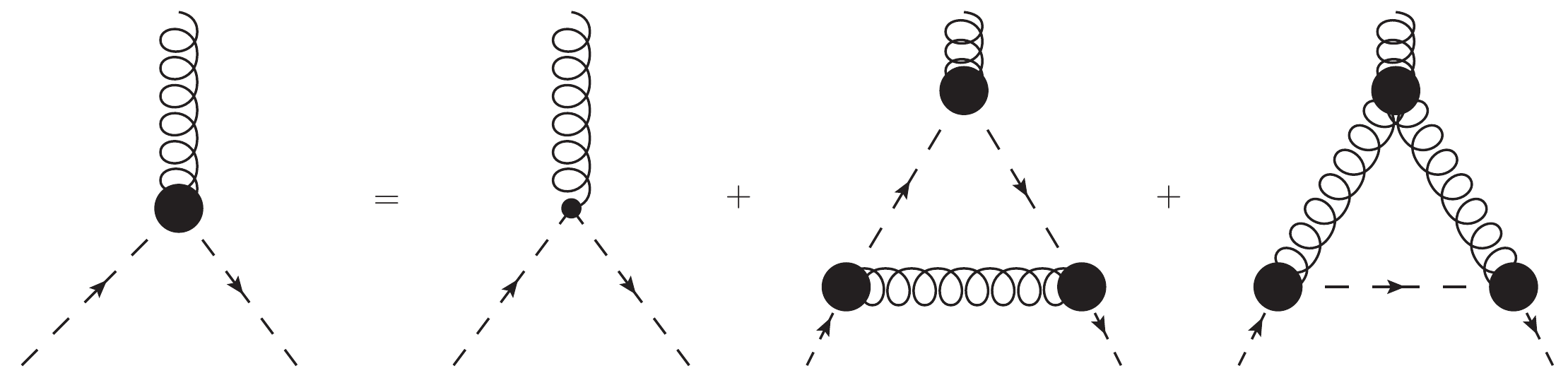}
    \caption{\label{fig:ghg-3PI}Equation of motion from the 3PI effective action for the ghost-gluon vertex.}
 \end{center}
\end{figure}

\begin{figure}[tb]
 \begin{center}
  \includegraphics[width=0.45\textwidth]{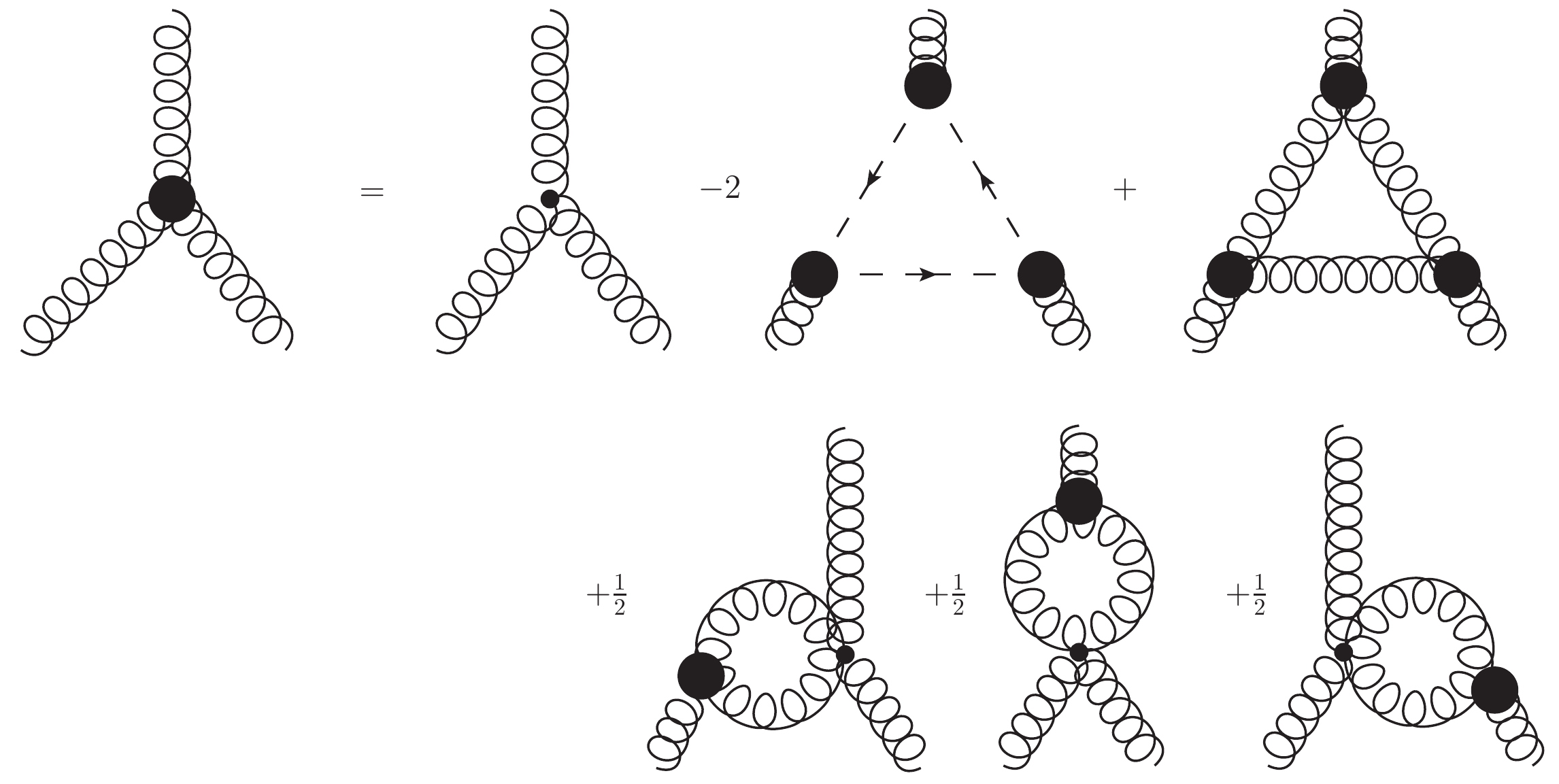}
  \caption{\label{fig:3g-3PI}Equation of motion from the 3PI effective action for the three-gluon vertex.}
 \end{center}
\end{figure}

As another set of equations the equations of motion from the 3PI effective action \cite{Berges:2004pu} are considered. The equations for the three-point functions are shown in Figs.~\ref{fig:ghg-3PI} and \ref{fig:3g-3PI}. The four-gluon vertex corresponds at the considered order to the bare vertex. The propagator equations, on the other hand, are identical to the DSEs in Figs.~\ref{fig:gh-DSE} and \ref{fig:gl-DSE}. Furthermore, it should be noted that the equations for the three-point functions  are very similar to the DSEs within the chosen truncation scheme and only differ in the numbers of dressed vertices.

In four dimensions the Yang-Mills system has been studied very intensively with DSEs; see, e.g., Ref.~\cite{Huber:2015fna} for a short overview. Although there are differences between the three- and the four-dimensional theories, as summarized in Sec.~\ref{sec:YM3d}, they have many common features. For example, both theories are asymptotically free and confining. Furthermore, evidence indicates that the functional structures of the two theories are very close so that we can learn something about four-dimensional Yang-Mills theory from the three-dimensional one. In particular, from lattice calculations it is known, and supported by this work, that the relevant qualitative behavior of correlation functions in the nonperturbative regime is the same \cite{Cucchieri:2003di,Cucchieri:2004mf,Cucchieri:2008qm,Maas:2008ri,Maas:2009se,Maas:2009ph,Cucchieri:2006tf,Cucchieri:2008fc,Cucchieri:2007rg,Cucchieri:2009zt,Cucchieri:1999sz,Cucchieri:2011ig,Bornyakov:2011fn,Maas:2010qw,Maas:2011se, Bornyakov:2013ysa,Maas:2014xma,Cucchieri:2016jwg}: The gluon propagator is nonvanishing at zero momentum and has a bump around $1\,\text{GeV}$; the ghost dressing function is finite; the ghost-gluon vertex is close to the tree-level; the three-gluon vertex has a zero crossing. As far as the hierarchy of diagrams is concerned, both the UV and the IR hierarchy are the same. For the former this statement is trivial, since it is the same loop expansion in the coupling. The latter case is strongly supported by the results obtained in this work which compare favorably with all available results in four dimensions even down to the hierarchy of diagrams in the gluon propagator DSE \cite{Meyers:2014iwa,Huber:2014tva}. For the deep IR and the scaling type solution it is even possible to show that the complete hierarchy of correlation functions, viz., without applying any truncations, is the same as in four dimensions \cite{Huber:2007kc}. Although the scaling solution is not seen in lattice calculations, it is a viable choice for functional equations and can, as in this case, provide additional information. In summary, all evidence supports that three- and four-dimensional Yang-Mills theories work similarly. Before turning to the three-dimensional theory, for reference the main results in four dimensions are summarized below. Some of them will be explicitly tested in three dimensions.

The importance of ghost contributions was recognized long ago \cite{vonSmekal:1997is,vonSmekal:1997vx} and subsequently often confirmed, e.g., \cite{Zwanziger:2001kw,Pawlowski:2003hq,Aguilar:2013vaa,Huber:2014tva}. Qualitative agreement with lattice results \cite{Cucchieri:2007md,Cucchieri:2008fc,Sternbeck:2007ug,Bogolubsky:2009dc} in the IR region was obtained when the so-called decoupling solution was found \cite{Boucaud:2008ji,Aguilar:2008xm,Fischer:2008uz}.
All the results from functional equations up to that point used models for the vertices. The ghost-gluon vertex was typically taken as bare, which is indeed comparatively close to the true vertex \cite{Cucchieri:2008qm,Schleifenbaum:2004id,Alkofer:2008dt,Huber:2012kd,Aguilar:2013xqa} although the deviations do have a quantitative effect on the propagators \cite{Huber:2012kd,Aguilar:2013xqa}. A three-gluon vertex model that contains the most notable features of the vertex like a zero crossing in the IR and Bose symmetry was introduced in Ref.~\cite{Huber:2012kd}. As expected, the three-gluon vertex has a severe quantitative, but not qualitative impact on the gluon propagator. When results for the three-point functions became available, it turned out that even with such improved input the gluon propagator cannot be described satisfactorily \cite{Blum:2014gna}. Consequently, the missing contributions were attributed to the two-loop diagrams, which were investigated previously in Refs.~\cite{Bloch:2003yu,Mader:2013ru,Meyers:2014iwa,Hopfer:2014th}. On the other hand, the found level of agreement between the available results for three-point functions and lattice results indicates that a one-loop truncation is satisfactory in this case \cite{Blum:2014gna,Eichmann:2014xya}.

The highest calculated dressing function up to now is the four-gluon vertex. Calculations with fixed input showed that the tree-level dressing function has only relatively small nonperturbative contributions \cite{Binosi:2014kka,Cyrol:2014kca}. The same is valid for further dressing functions investigated in Ref.~\cite{Cyrol:2014kca}. The four-gluon vertex appears in the so-called sunset diagram of the gluon propagator and is thus required for the direct calculation of two-loop diagram effects.

As explained, in four dimensions all primitively divergent Green functions have already been calculated, partly even from coupled systems. It should be stressed that retaining only these vertices in all diagrams, the resulting truncated set of DSEs is self-contained. At least in the three-dimensional case it seems also to be the lowest working nontrivial truncation without models, since lower truncations, for example, with the propagators and the ghost-gluon vertex included and all other vertices set to zero, did not lead to convergent solutions. Furthermore, the four-gluon vertex is required to have a consistent treatment of the UV behavior which is a minimum requirement for a truncation working well at all momenta. It remains to be tested if the calculation of all five primitively divergent Green functions indeed provides a good quantitative description.

One crucial point in answering this question concerns the treatment of spurious divergences which plague the gluon propagator DSE. Looking at the four-dimensional case, power counting shows that the superficial degree of divergence is 2; viz., the equation is quadratically divergent. However, gauge symmetry entails that the gluon self-energy is proportional to $g_{\mu\nu}p^2-p_\mu p_\nu$, which lowers the degree of divergence by 2, and it boils down to a logarithmic divergence. Unfortunately, many regularizations break gauge symmetry and quadratic divergences reappear. This is not a problem specific to DSEs but occurs for such regularizations already at the perturbative level. Since these divergences do not appear in dimensional regularization, which is the standard regularization in perturbation theory, they are not problematic for many perturbative investigations, though. For a numerical calculation dimensional regularization is at best difficult to realize and maybe not even applicable at all \cite{Gusynin:1998se,Phillips:1999bf}. The standard numerical regularization is a hard UV cutoff for which quadratic divergences that have to be dealt with appear.

In the literature many ways can be found to reduce the cutoff dependence to a logarithmic one, see Ref.~\cite{Huber:2014tva} for an overview. However, this procedure is not unique due to the finite part of the divergences. Unfortunately, the IR leading part of the gluon propagator DSE has the same $1/p^2$ behavior as the spurious divergences and the separation of the infinite and the finite parts becomes crucial. In Ref.~\cite{Huber:2014tva} it was shown that the origin of spurious divergences is purely perturbative. Thus, one necessary requirement of any subtraction procedure should be that it does not interfere with the nonperturbative part. This restricts the number of choices for the subtraction of spurious divergences.

Here I extend the approach put forward in Ref.~\cite{Huber:2014tva}, which is a minimal subtraction where the subtracted terms correspond exactly to those generated in perturbation theory. In the case of a truncation that only involves the propagators dynamically and uses models for the vertices, the subtraction terms can be calculated analytically. As soon as the vertices are treated dynamically as well, this is no longer possible. First of all, the perturbative expression with full momentum dependence in cutoff regularization, which is required for such a calculation, is not known analytically for the three-gluon vertex. Second, it is not clear yet how well the UV behavior at the one-loop resummed level, which is also required for this calculation, can be reproduced for the three-gluon vertex \cite{Blum:2014gna,Eichmann:2014xya}. Even for the ghost-gluon vertex, which is asymptotically trivial, the small corrections to the tree-level behavior at finite momenta are important and spoil the naive approach of taking for such calculations a trivial vertex in the UV. For two-loop diagrams the same problems apply.

In three dimensions the case is simpler. The main reason for this is the absence of renormalization (except for the spurious divergences). This makes the UV behavior of the dressings simpler. The procedure is explained in Appendix~\ref{sec:spurDivs}. To summarize it, two-loop diagrams and vertices can be taken into account by fitting the subtraction coefficient to the known analytic form.

\section{Yang-Mills theory in three dimensions}
\label{sec:YM3d}

Going from four to three space-time dimensions has several advantages, while it is expected that many qualitative features of Green functions remain the same; see, e.g., Ref.~\cite{Teper:1998te}. However, the advantages are distinct for continuum and lattice methods. For the latter, the reduction of space-time dimensions allows larger lattices to be used, and thus the IR regime is more easily accessible. Furthermore, the reduced computing requirements allow better statistics and thus make three dimensions also interesting for the investigation of the Gribov problem; see, e.g, Ref.~\cite{Maas:2015nva} and references therein. Propagators have been studied quite extensively \cite{Cucchieri:2003di,Cucchieri:2004mf,Cucchieri:2008qm,Maas:2008ri,Maas:2009se,Maas:2009ph,Cucchieri:2006tf,Cucchieri:2008fc,Cucchieri:2007rg,Cucchieri:2009zt,Cucchieri:1999sz,Cucchieri:2011ig,Bornyakov:2011fn,Maas:2010qw,Maas:2011se, Bornyakov:2013ysa,Maas:2014xma,Cucchieri:2016jwg}, whereas for three-point functions results are similarly as scarce as in four dimensions \cite{Cucchieri:2006tf,Cucchieri:2008qm,Maas:2016ip}. However, the statistics are better for three than for four dimensions.

\begin{figure}[tb]
 \includegraphics[width=0.45\textwidth]{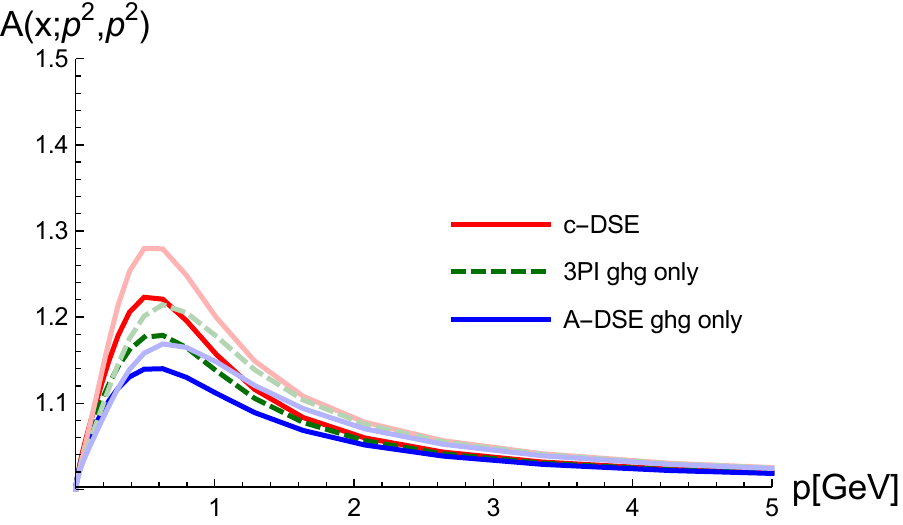}
 \hfill
\caption{\label{fig:ghgCompEqs}Ghost-gluon vertex dressing function calculated with fixed input from different equations. Dark/light lines correspond to $x=0$/$x=p^2$.} 
\end{figure}

The lower number of space-time dimensions does not directly affect the required computing time for functional equations of low n-point functions since some integrations can be done analytically. Only for four-point functions one integration fewer has to be done and computing time is reduced. The reason for using three dimensions in this work is the absence of renormalization and thus resummation that entails a significant simplification of the treatment of spurious divergences. In two dimensions similar arguments apply, but already standard perturbation theory fails in two dimensions due to IR singularities. Also for DSE calculations it was found that the mixing of the IR and UV regimes leads to nontrivial UV problems \cite{Huber:2012zj} which make two-dimensional Yang-Mills theory less suitable for the present investigation.

A physical motivation for understanding three-dimensional Yang-Mills theory is, upon adding an adjoint scalar, that it is the high temperature asymptotic limit of the four-dimensional theory. This system was investigated in Ref.~\cite{Maas:2004se}. Further continuum studies of three-dimensional Yang-Mills theory can be found in Refs.\cite{Maas:2004se,Huber:2007kc,Alkofer:2008dt,Aguilar:2010zx,Aguilar:2013vaa,Cornwall:2015lna}. Within the (refined) Gribov-Zwanziger (GZ/RGZ) framework it was investigated in Ref.~\cite{Dudal:2008rm}. In Refs.~\cite{Tissier:2010ts,Tissier:2011ey,Pelaez:2013cpa} a massive extension of Yang-Mills theory was considered.

One consequence of lowering the dimension is that all equations become finite with the exception of the gluon DSE where spurious divergences appear. The lower dimension is also reflected in the coupling constant $g$ that has the dimension of $(\text{mass})^\frac{1}{2}$. Furthermore, the lower dimension of the integral has some direct consequences for the asymptotic regimes. In the UV, the dressings behave at leading order like $1/p$; see Appendix~\ref{sec:spurDivs}. In the IR, on the other hand, logarithmic divergences as appear, for instance, in the vertices \cite{Pelaez:2013cpa,Aguilar:2013vaa,Blum:2014gna,Eichmann:2014xya} typically become linear. For the three-gluon vertex this was already found in Refs.~\cite{Pelaez:2013cpa,Aguilar:2013vaa}. Here this is confirmed and observed for the four-gluon vertex as well.

\begin{figure}[tb]
 \includegraphics[width=0.45\textwidth]{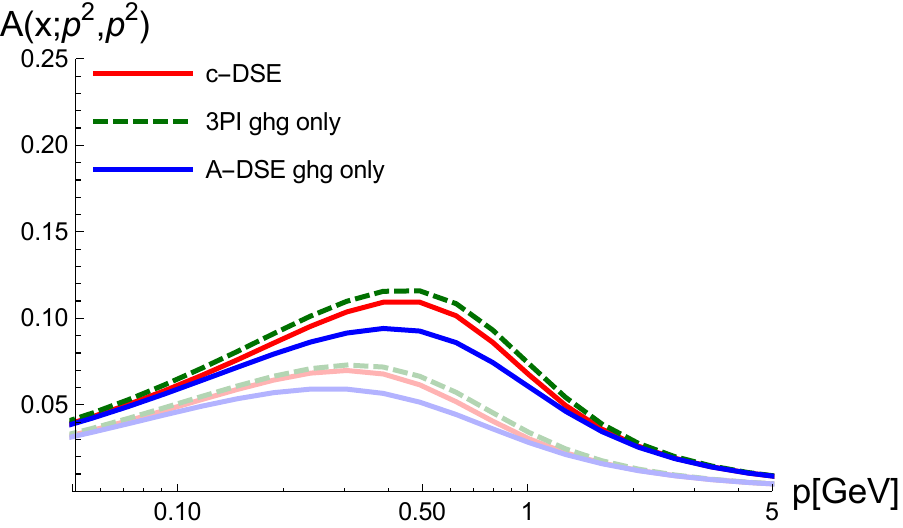}\\
 \vskip2mm
 \includegraphics[width=0.45\textwidth]{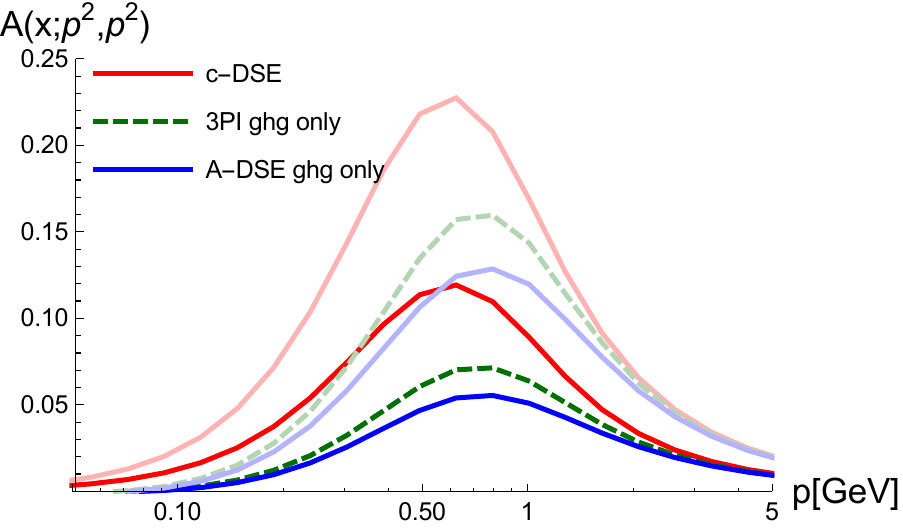}
\caption{\label{fig:ghgCompEqs-diags}Contributions of the Abelian (top) and non-Abelian (bottom) diagrams of the ghost-gluon vertex DSE calculated with fixed input from different equations. Dark/light lines corresponds to $x=0$/$x=p^2$.} 
\end{figure}

\section{Choice of equations}
\label{sec:ghg}

In this section two comparisons are done. First, results from the three different equations for the ghost-gluon vertex are compared. Second, results from the full DSE truncation are compared to results from the 3PI effective action.

\subsection{Ghost-gluon vertex equation}

In the employed truncation scheme only two diagrams remain in the ghost-gluon vertex equations, the Abelian and non-Abelian triangles. The only difference between the different equations consists in the position of the bare vertex or its nonexistence in case of the 3PI equation. In case of the $c$-DSE, only one diagram containing a ghost-gluon four-point function is dropped, while for the $A$-DSE nine diagrams are dropped (seven two-loop diagrams and two one-loop diagrams containing a quartic ghost function or a ghost-gluon four-point function; see Ref.~\cite{Alkofer:2008nt} for the full equations). A priori there is no reason to expect that different equations, once they are truncated, yield quantitatively equal results. The difference between solutions can be interpreted as an estimate of the truncation error. From the ghost-gluon vertex example it is also obvious that, depending on the truncation, one equation might be the better choice. For instance, if the ghost-gluon four-point function is added, the $c$-DSE becomes an exact equation, while in the $A$-DSE there are still missing diagrams. Within the current truncation, however, the contribution of the dropped diagram in the $c$-DSE is not known and we do not know which equation yields better results.

To compare results from the three different equations, they are solved with fixed input that was obtained from the full system; see Sec.~\ref{sec:results}. In \fref{fig:ghgCompEqs} the obtained dressings of the ghost-gluon vertex are shown. The contributions from the single diagrams are depicted in \fref{fig:ghgCompEqs-diags}. As can clearly be seen, the height of the vertex dressing varies with the $A$-DSE yielding the lowest and the $c$-DSE the highest dressing. The main difference comes, as expected, from the non-Abelian diagram. Interestingly, the result from the 3PI effective action is very close to the average of the two DSE results. Using the maximal ratio of the maxima of the complete set of points obtained for each equation as an estimate of the truncation error $e_\text{ghg}$, it can be quantified as $e_\text{ghg}=13\,\%$. Note that the difference for the configurations shown is even lower and typically below $10\,\%$.

\begin{figure}[tb]
 \includegraphics[width=0.45\textwidth]{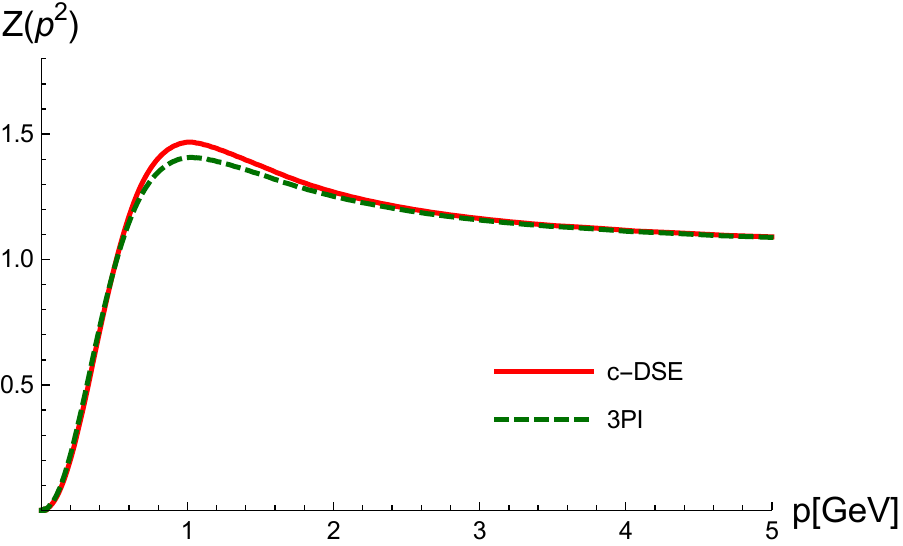}
 \caption{\label{fig:glCompDSE3PI}Comparison of results from the DSE and the 3PI systems for the gluon dressing function.} 
\end{figure}

\subsection{DSEs vs. 3PI}

\begin{figure}[tb]
 \includegraphics[width=0.45\textwidth]{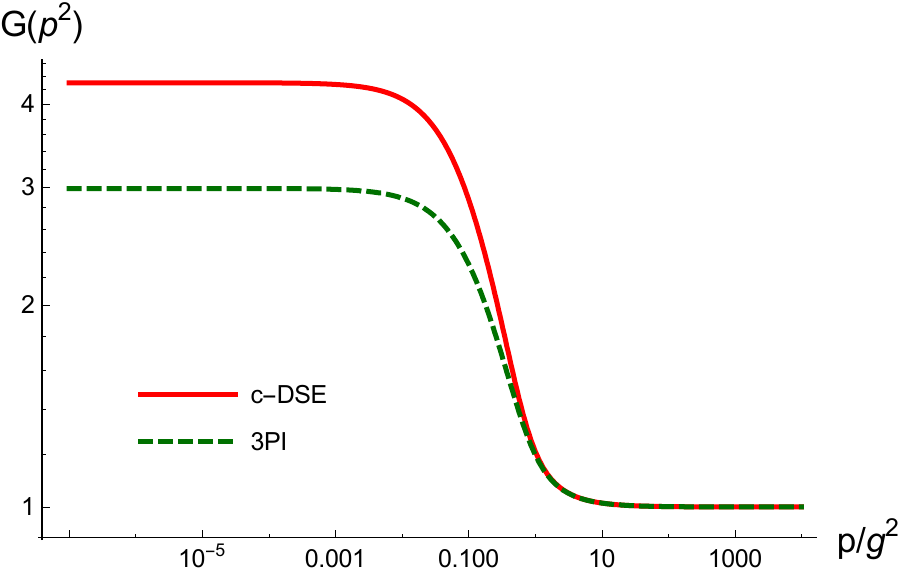}
 \caption{\label{fig:ghCompDSE3PI}Comparison of results from the DSE and the 3PI systems for the ghost dressing function.} 
\end{figure}

\begin{figure*}[tb]
 \includegraphics[width=0.45\textwidth]{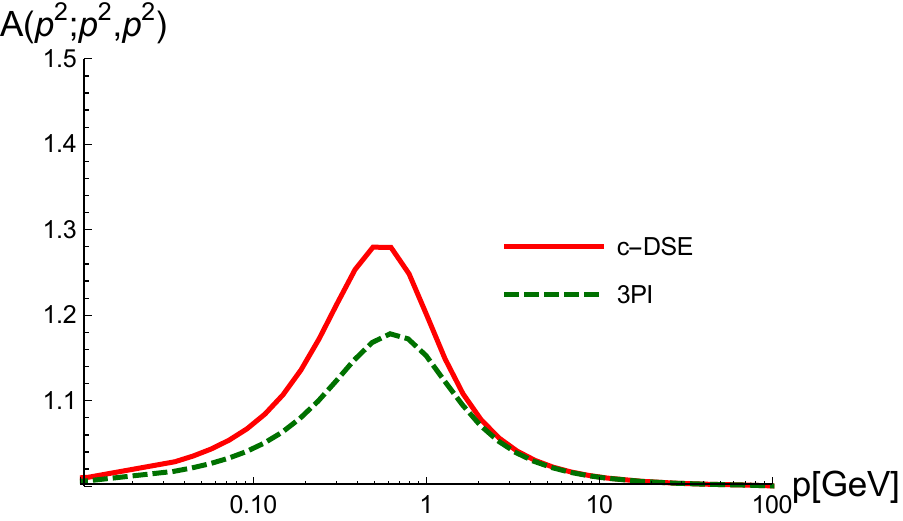}
 \hfill
 \includegraphics[width=0.45\textwidth]{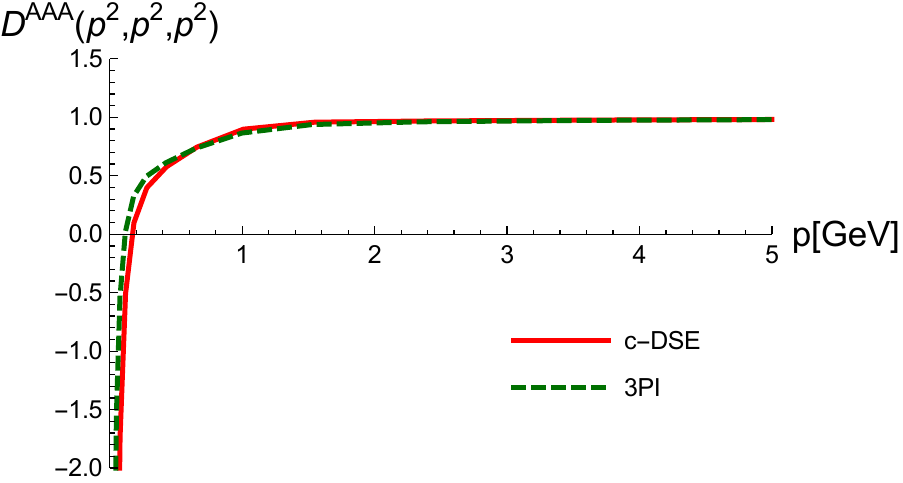}
 \caption{\label{fig:ghgTgCompDSE3PI}Comparison of results from the DSE and the 3PI systems for the ghost-gluon vertex and three-gluon vertex dressing functions.} 
\end{figure*}

To assess the influence the uncertainty in the ghost-gluon vertex has on other correlation functions, the full systems of equations are now considered for DSEs and the 3PI effective action. For the DSE system the $A$-DSE is chosen for the ghost-gluon vertex. The corresponding results are depicted in Figs.~\ref{fig:glCompDSE3PI}, \ref{fig:ghCompDSE3PI} and \ref{fig:ghgTgCompDSE3PI}. The error estimation from the ghost-gluon vertex is $e_\text{ghg}=12\,\%$ with the 3PI results being below the results from the DSE as can be seen in \fref{fig:ghgTgCompDSE3PI}. The fact that the ghost-gluon vertex dressing from the 3PI effective action is reduced is also known from four dimensions \cite{Williams:2015cvx}. The effect in the three-gluon vertex, shown in \fref{fig:ghgTgCompDSE3PI}, is very small with a shift of the zero crossing the most notable change; viz., changes are only visible below $500\,\text{MeV}$. The propagators, see \fref{fig:glCompDSE3PI}, also differ in the IR, while a difference in the midmomentum regime is only visible for the gluon. In summary, both DSEs and 3PI equations yield similar results with the largest differences for the ghost-gluon vertex in the midmomentum regime and for the propagators in the deep IR. Given the found degree of agreement, one can conclude that using DSEs or 3PI equations leads to very similar results. Since in the 3PI formalism no dressed four-gluon vertex appears within the employed truncation, this setup is technically easier.

\section{Higher Green functions: Effect of the four-gluon vertex}
\label{sec:4g_effects}

\begin{figure}[tb]
 \includegraphics[width=0.48\textwidth]{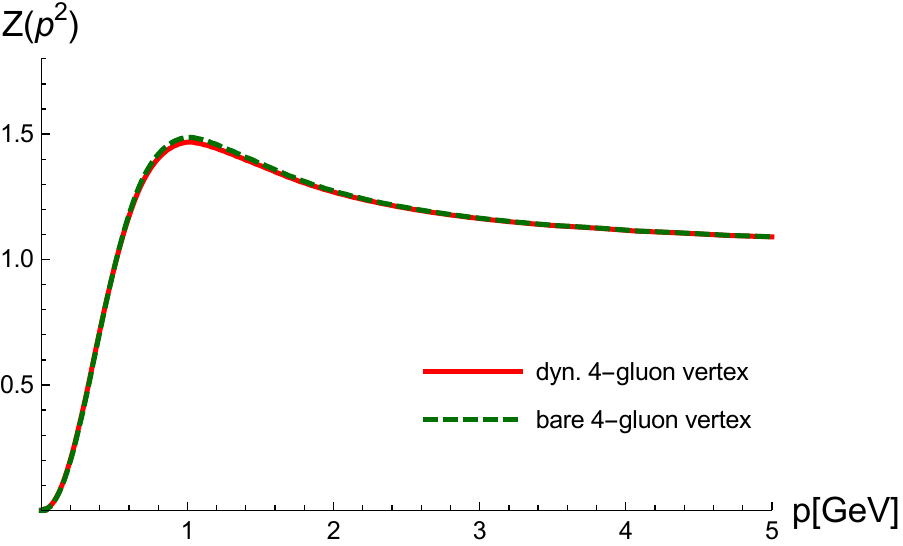}
 \caption{\label{fig:glComp4gs}Gluon dressing function from the full system with a bare (green, dashed line) and a dynamic four-gluon vertex (red, continuous line).}
\end{figure}

\begin{figure}[tb]
 \includegraphics[width=0.48\textwidth]{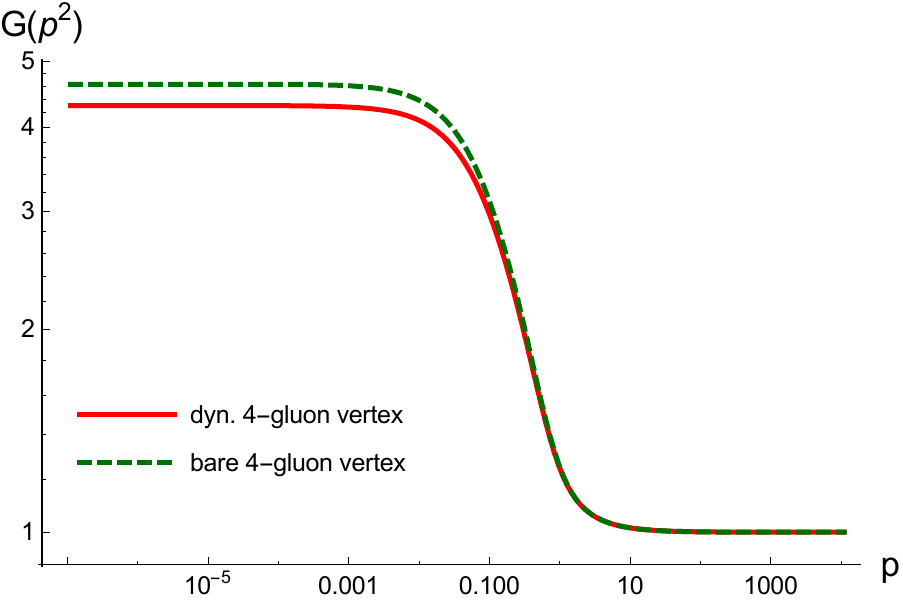}
 \caption{\label{fig:ghComp4gs}Ghost dressing from the full system with a bare (green, dashed line) and a dynamic four-gluon vertex (red, continuous line).}
\end{figure}

\begin{figure}[tb]
 \includegraphics[width=0.48\textwidth]{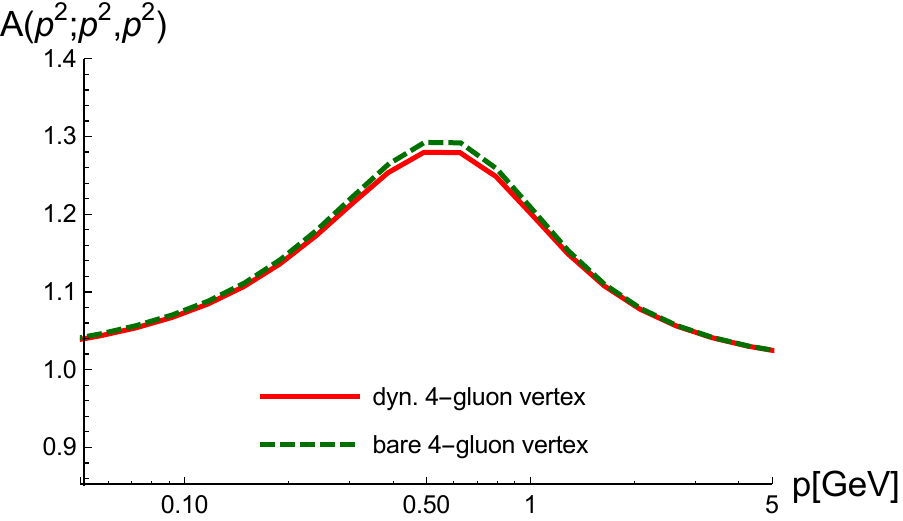}
 \vskip2mm
 \includegraphics[width=0.48\textwidth]{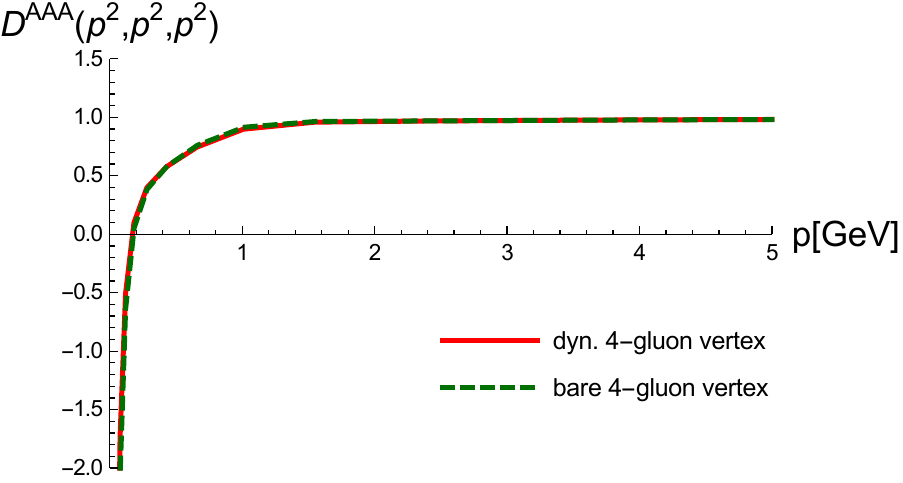}
 \caption{\label{fig:ghgTgComp4gs}Ghost-gluon vertex (top) and three-gluon vertex (bottom) dressing from the full system with a bare (green, dashed line) and a dynamic four-gluon vertex (red, continuous line).}
\end{figure}

One possibility to test the reliability of a truncation is to include more quantities dynamically. However, such tests must be interpreted with care. The reason is twofold. First of all, one has to keep in mind what the function of the model that is replaced by its dynamic counterpart was. In case it was designed to improve the results for some other quantity without regard to its correct behavior, this replacement might not improve the results. A typical example is the three-gluon vertex in the gluon propagator DSE. The choice of the model has a large impact on the quantitative behavior of the gluon dressing function. In fact, models exist that can be tuned such that the gluon propagator is in good agreement with lattice results thereby effectively mimicking two-loop effects \cite{Huber:2012kd}. This also works when quarks are included \cite{Williams:2015cvx}. When this model is compared to results from the lattice \cite{Cucchieri:2008qm} and dynamical DSE calculations, clear differences in the nonperturbative regime are visible \cite{Huber:2012kd,Aguilar:2013vaa,Blum:2014gna,Eichmann:2014xya}. In this case, adding the three-gluon vertex dynamically requires the inclusion of other quantities as well in order to obtain good results.

The second caveat is the quality of the truncation of the dynamical equation for the added quantity. While this seems like having the same problem again at the next level, it turns out at least in Yang-Mills theory that for higher Green function truncation effects become less severe. This is indicated by results for three-point functions, which compare favorably to lattice results \cite{Huber:2012kd,Aguilar:2013xqa,Blum:2014gna,Eichmann:2014xya}, as well as by the observation that the deviation from the tree-level becomes small for the four-gluon vertex \cite{Binosi:2014kka,Cyrol:2014kca}. In this work this is confirmed in Sec.~\ref{sec:results}.

\begin{figure*}[tb]
 \includegraphics[width=0.48\textwidth]{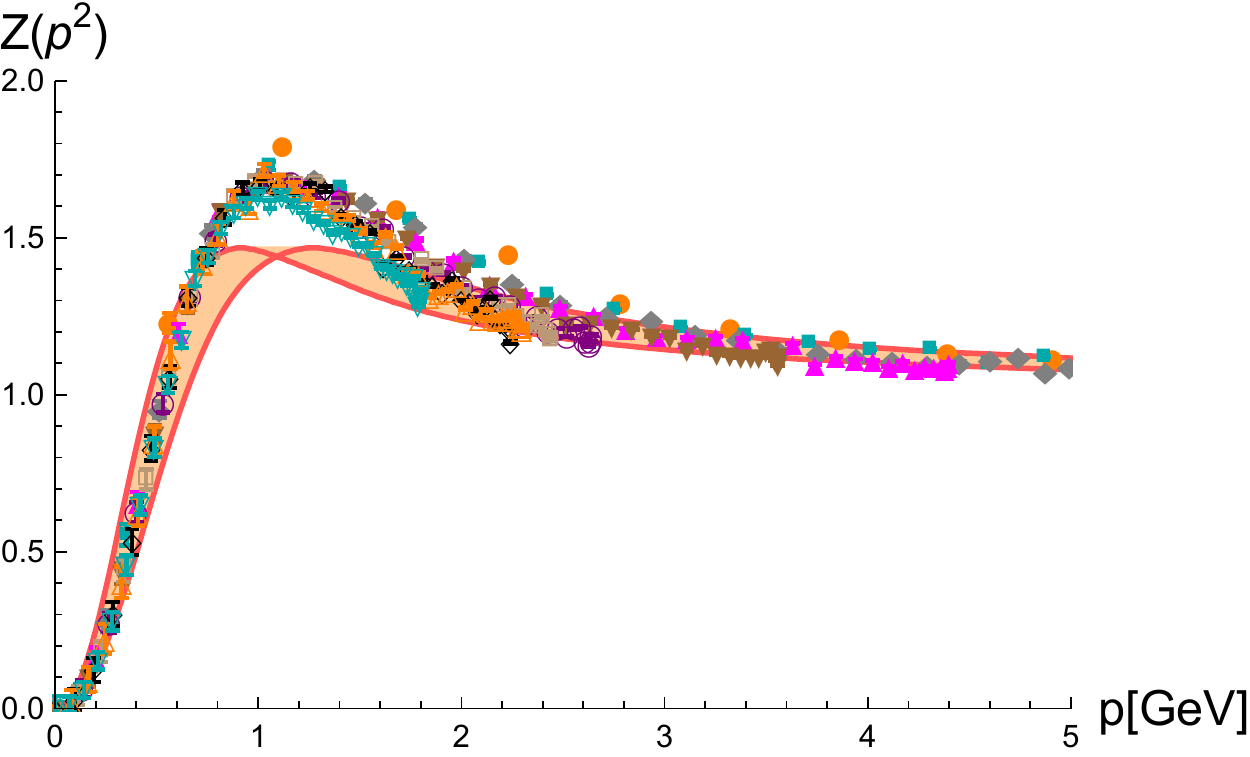}
 \hfill
 \includegraphics[width=0.48\textwidth]{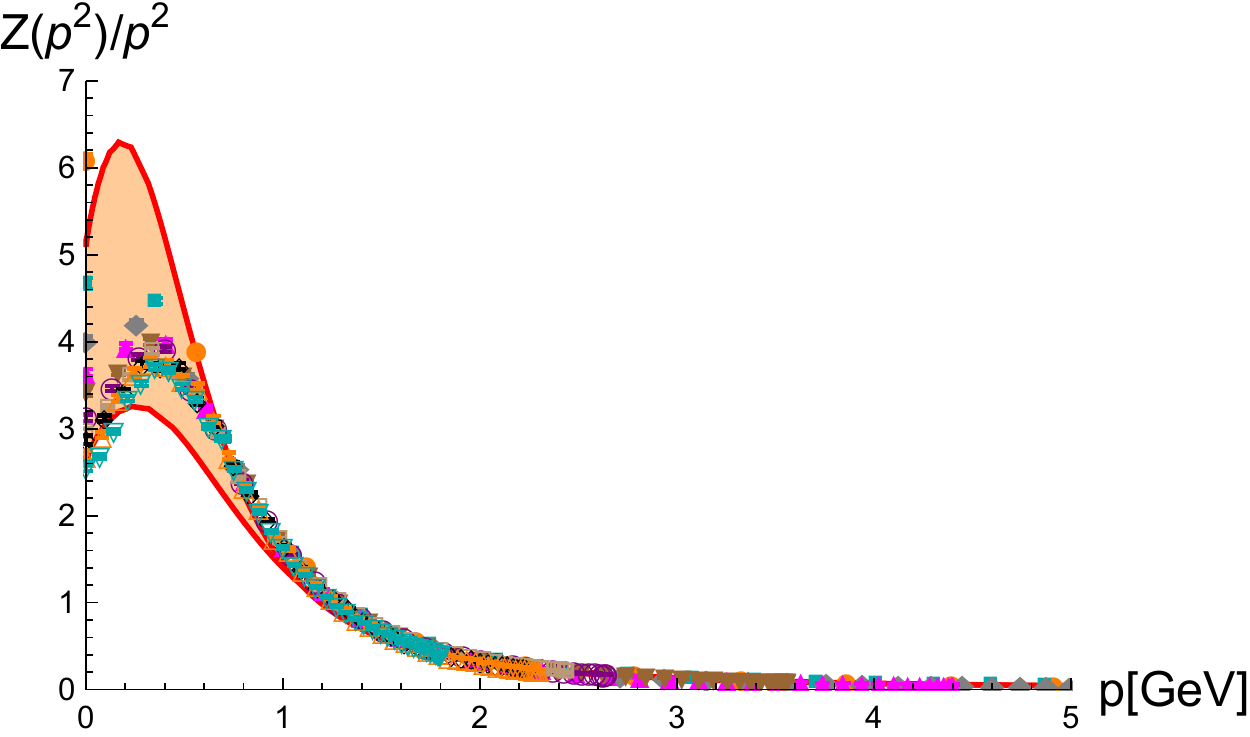}
 \caption{\label{fig:glWLattice}Gluon dressing function (left) and gluon propagator (right) from the full system in comparison to lattice results \cite{Maas:2014xma}. Lattice momenta are along the $x$ axis. The shown lattice results correspond to lattice sizes between $N=68$ and $N=88$ and $\beta$ values between $3.48$ and $19.2$. The band is obtained by varying the maximum of the gluon dressing function between $922$ and $1282\,\text{MeV}$.}
\end{figure*}

In this section we have a look at the impact of the four-gluon vertex as the highest Green function contained in the truncation. As will be shown in Sec.~\ref{sec:results}, see in particular \fref{fig:fg_V1}, the tree-level dressing function deviates in the midmomentum regime only mildly from the tree-level and an IR divergence is observed in the deep IR. To assess the influence of these deviations, the system of propagators and three-point functions was calculated using a bare four-point function. The results for the propagator dressing functions are shown in Figs.~\ref{fig:glComp4gs} and \ref{fig:ghComp4gs}, and those for the three-point functions in \fref{fig:ghgTgComp4gs}. It can clearly be seen that the difference is relatively small and a bare four-gluon vertex provides already a rather good approximation. However, it has to be pointed out that in four dimensions the model employed for the four-gluon vertex plays a crucial role in the three-gluon vertex DSE to obtain a convergent solution \cite{Blum:2014gna,Eichmann:2014xya}.

\section{Comparison with lattice results}
\label{sec:results}

In this section the results from the full system are compared to lattice results. Before making such comparisons, two things have to be discussed. The first one is related to the physical scale of the results from the functional equations. As no appropriate observable is calculated to determine the scale, it is inferred from the lattice results where the scale was set from the string tension with $\sigma=(440\,\text{MeV})^2$. The employed prescription is to set the maximum of the gluon dressing function at the same location as in the lattice results. However, since the DSE results are not on top of the lattice results, this procedure is ambiguous. Thus, when comparing directly to lattice data, a band is shown the boundaries of which are determined by assuming a variation of the maximum between $90$ and $125\,\%$. In \fref{fig:glWLattice} it can be seen that this corresponds to a plausible interval for the maximum. For the results shown in the previous sections, the maximum was set at $1.025\,\text{GeV}$.

\begin{figure}[tb]
 \includegraphics[width=0.48\textwidth]{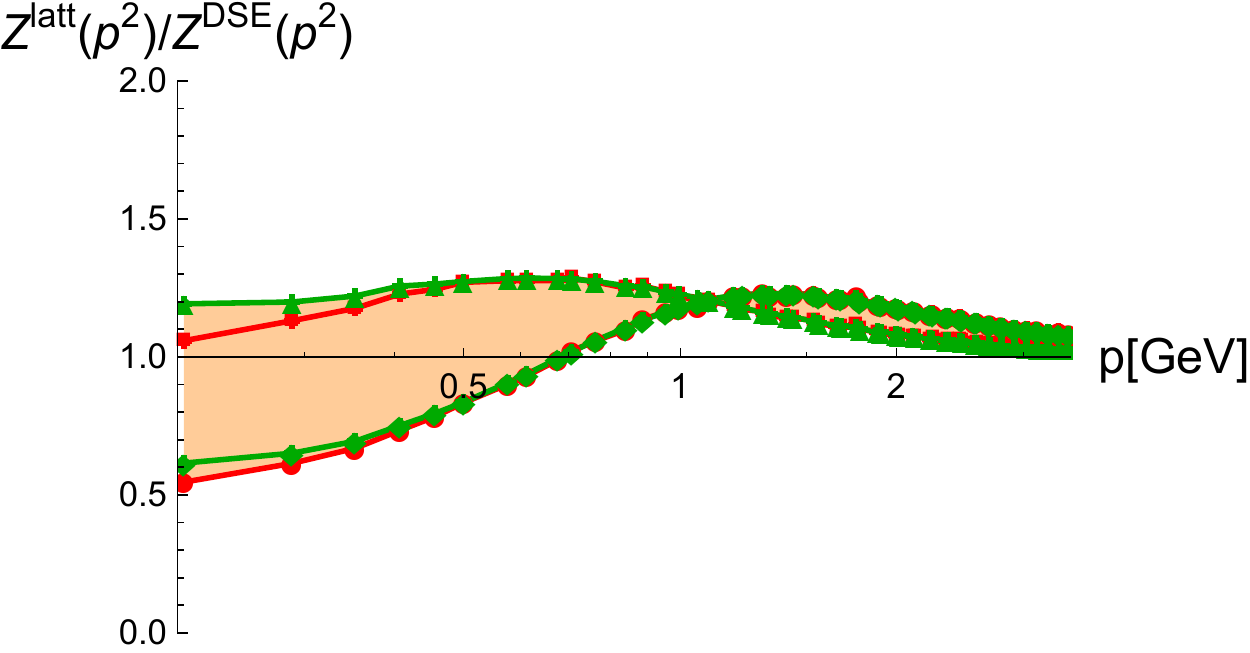}
 \caption{\label{fig:glRatio}Ratio of the gluon propagator from the lattice \cite{Bornyakov:2013ysa} and the full DSE system. Lattice data is for $\beta=10.21$ and $N=96$. The band is obtained by varying the maximum of the gluon dressing function between $922$ and $1282\,\text{MeV}$. The red (lower) and green (upper) branches of the lattice data correspond to using the best and the first Gribov copy in the gauge fixing algorithm, respectively.}
\end{figure}

\begin{figure}[tb]
 \includegraphics[width=0.48\textwidth]{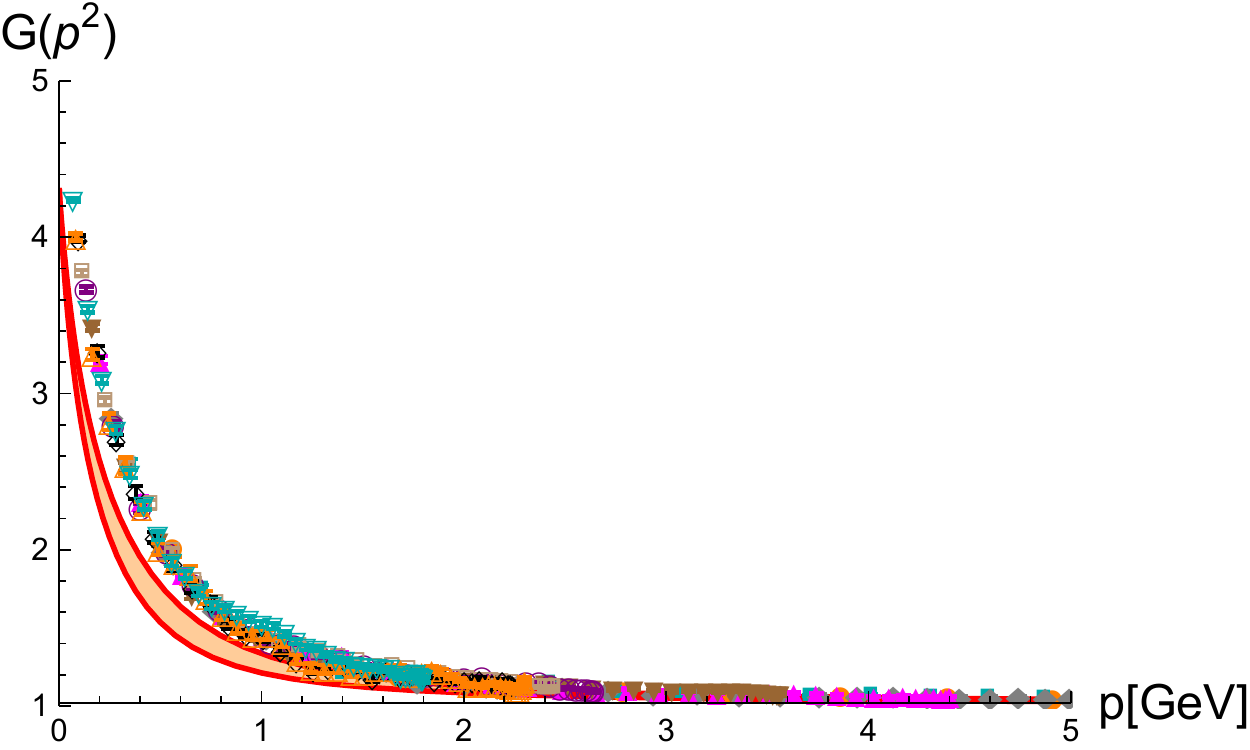}
 \caption{\label{fig:ghWLattice}Ghost dressing from the full system in comparison to lattice results \cite{Maas:2014xma}. The parameters of the lattice setup are the same as in \fref{fig:glWLattice}.}
\end{figure}

\begin{figure}[tb]
 \includegraphics[width=0.48\textwidth]{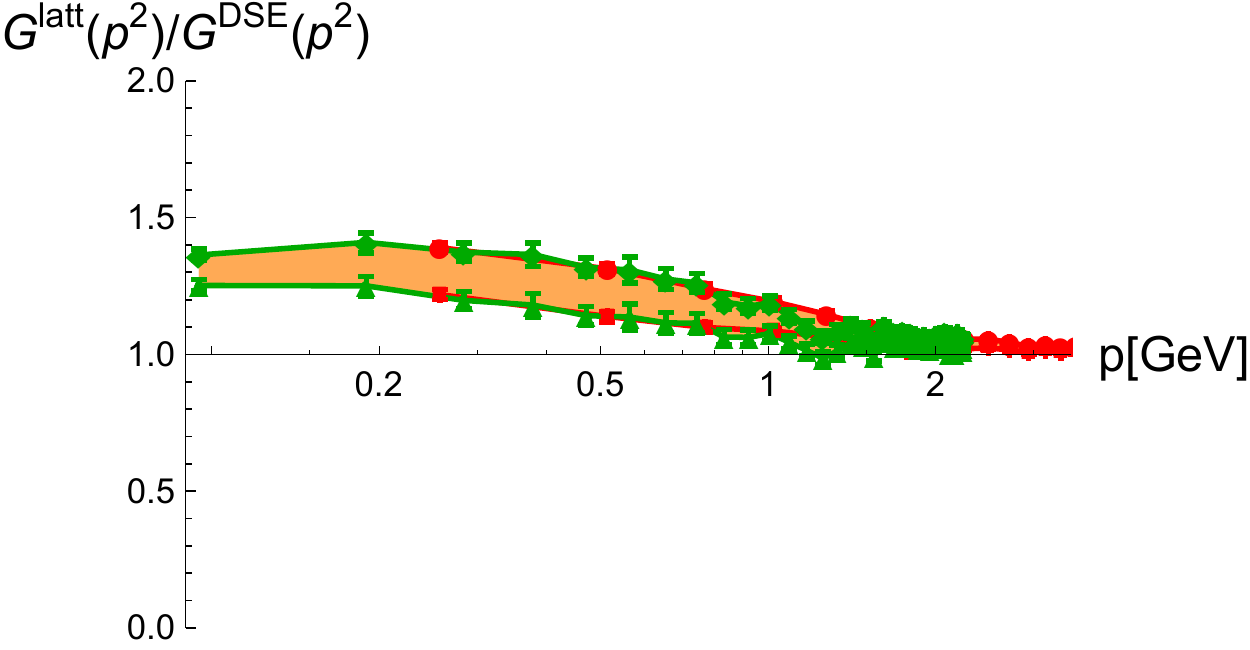}
 \caption{\label{fig:ghRatio}Ratio of the ghost propagator from the lattice \cite{Maas:2014xma} and the full DSE system. The parameters of the lattice setup are $N=68$ with $\beta=9.21$ and $N=74$ with $\beta=4.15$.}
\end{figure}

Another issue of comparisons between lattice and continuum results is related to the question of nonperturbative gauge fixing. There is evidence from lattice calculations that in the IR regime dressing functions depend on the gauge fixing algorithm; see, e.g., Refs.~\cite{Cucchieri:1997dx,Bogolubsky:2005wf,Sternbeck:2006rd,Maas:2009se,Maas:2011se,Sternbeck:2012mf} for four dimensions and \cite{Maas:2009se,Bornyakov:2011fn,Maas:2011se,Maas:2013vd,Bornyakov:2013ysa} for three dimensions. Also, for functional equations a family of solutions exists in four dimensions \cite{Boucaud:2008ji,Fischer:2008uz,Alkofer:2008jy}. Unfortunately, it is not clear how to set solutions from the lattice, where the differences come from choosing different but physically equivalent gauge configurations, and from continuum calculations, where different members of this family are chosen via a boundary condition of the ghost propagator DSE \cite{Fischer:2008uz}, into a direct relation.

To illustrate the magnitude of such effects, it is instructive to look at results for the ghost dressing function from Ref.~\cite{Maas:2013vd} where different sampling procedures are used. The two most extreme results for the ghost dressing function differ at $500\,\text{MeV}$ by $50\,\%$ and more at lower momenta. The lower solution even shows a maximum. Up to now such effects have been investigated only for propagators, but they are likely to exist also for vertices to cancel any effects on physical observables.

In this work no specific value for the ghost boundary condition was chosen as, in contrast to typical calculations in four dimensions, the unsubtracted ghost propagator DSE was used. When using the subtracted equation, the boundary condition could be set to a specific value, but the ghost dressing function was found not to connect smoothly to the UV regime. In four dimensions this works straightforwardly because of the logarithmic running of the dressing functions in the UV. However, it appears that some more elaborate technique is necessary to produce a family of solutions in three dimensions. This is also supported by the fact that the vertices have a significant influence on the specific IR behavior as was explicitly investigated for the system of the propagators alone using models for the vertices. For example, with one particular class of models for the three-gluon vertex it was even possible to obtain a scaling solution using the unsubtracted ghost equation by only demanding the appropriate IR extrapolation.
Since it is not known how close the lattice solution should be to the solution obtained here, the presented comparisons can only give some impression about the agreement. In the plots results from the minimal Landau gauge are used except for \fref{fig:glRatio} where also the result for the absolute Landau gauge is shown.

In \fref{fig:glWLattice} the gluon dressing function and the gluon propagator are shown. Due to the factor $1/p^2$ compared to the dressing function the differences for the propagator are most pronounced at low momenta. To explicitly show the magnitude of the deviation, the ratio of the gluon propagator obtained from the lattice over the result from DSEs is shown in \fref{fig:glRatio}. In addition to the band obtained from varying the scale, two different solutions from the lattice that vary only in the way the gauge is fixed are shown . In one case the first Gribov copy is taken, whereas in the other the gauge fixing algorithm chooses the Gribov copy with the lowest integrated gluon propagator. Below $500\,\text{MeV}$ these two algorithms yield different results which survive also in the continuum limit~\cite{Bornyakov:2013ysa}.

\begin{figure}[tb]
 \includegraphics[width=0.48\textwidth]{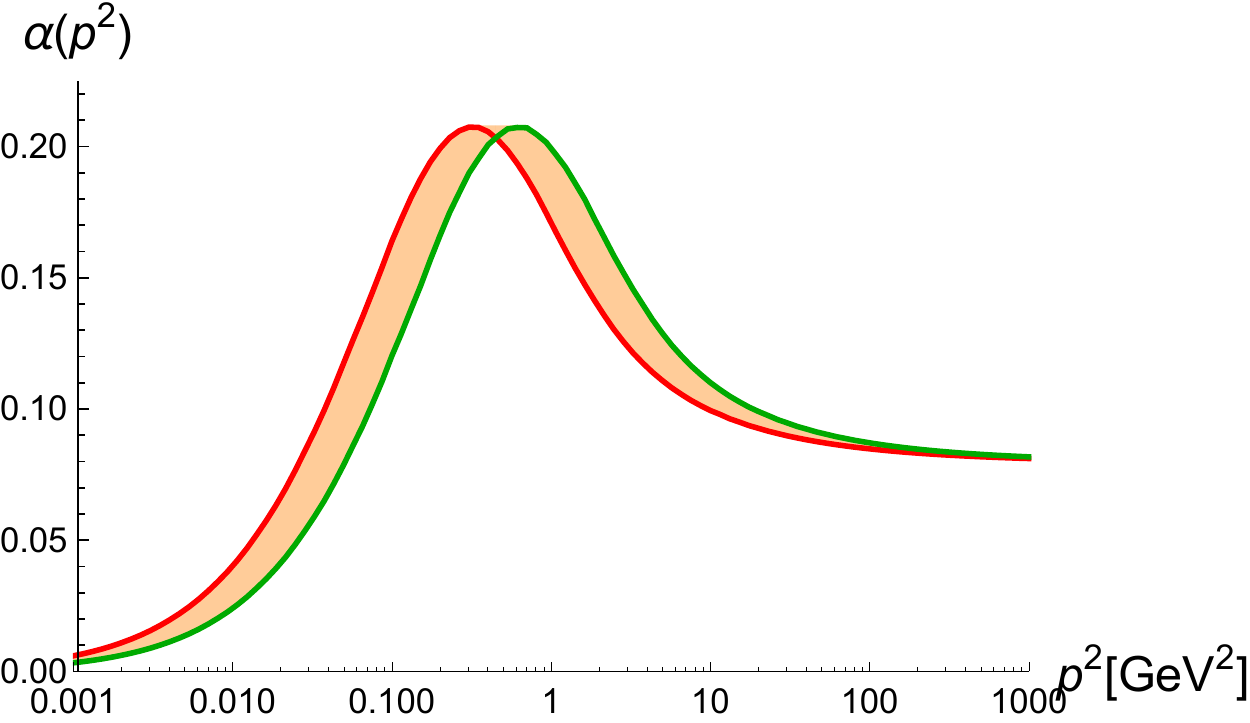}
 \caption{\label{fig:coupling}Coupling calculated from the propagators.}
\end{figure}

The ghost propagator dressing function is shown in \fref{fig:ghWLattice}, and the ratio between lattice and continuum results is shown in \fref{fig:ghRatio}. The ghost propagator from the DSE is systematically below the lattice results. However, the extreme solution mentioned above with a maximum of the dressing function \cite{Maas:2013vd} lies in the IR below the continuum solution. The coupling, calculated from the propagator dressing function as
\begin{align}
 \alpha(p^2)=\frac{g^2}{4\pi}G^2(p^2)Z(p^2),
\end{align}
is shown in \fref{fig:coupling}.

The behavior of both propagators follows in the UV the one-loop form, $1+c_\text{gh/gl}g^2/p$, but the coefficients $c_\text{gh}$ and $c_\text{gl}$ are slightly modified at the level of $5\,\%$. See Appendix~\ref{sec:spurDivs} for the exact perturbative expressions. This was already observed in Ref.~\cite{Maas:2014xma}, where it was found that the lattice data for the ghost and the gluon dressing functions cannot be described with the same value for the coupling $g$. Possible sources of this modification are nonperturbative contributions or higher order corrections. 

\begin{figure}[tb]
 \includegraphics[width=0.48\textwidth]{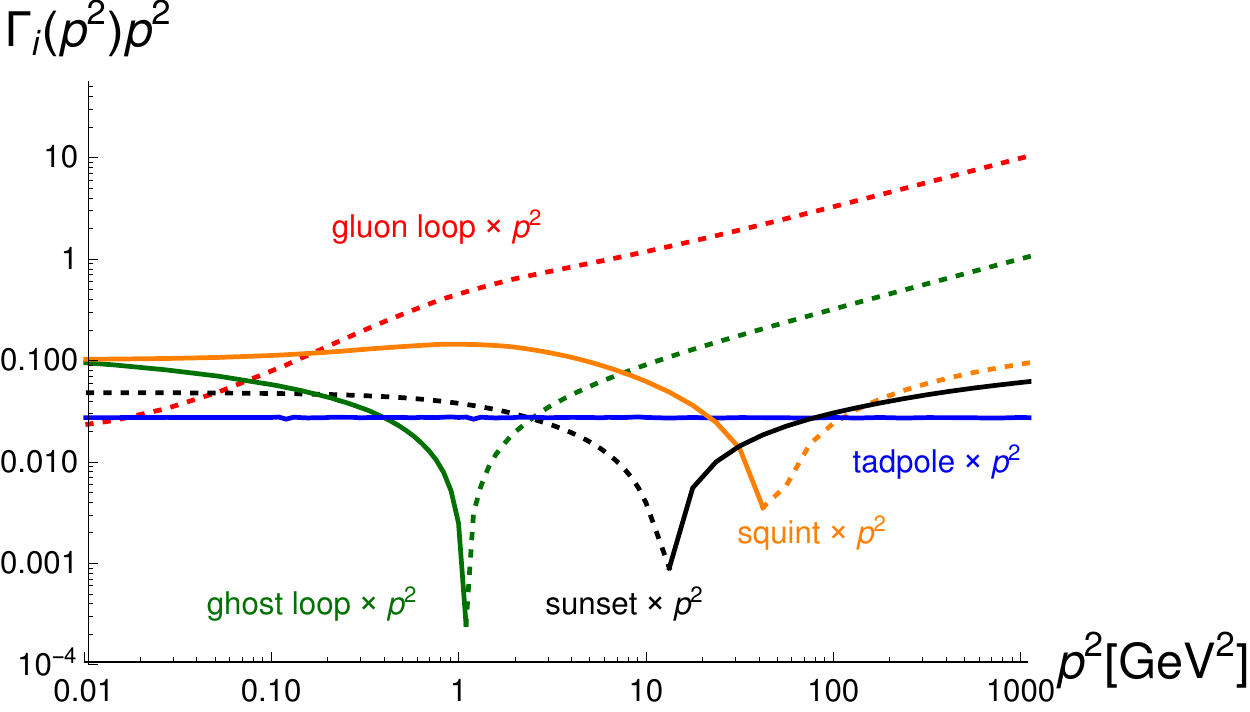}
 \caption{\label{fig:glDiagrams}Contributions of individual diagrams to the gluon propagator DSE. Continuous/dotted lines denote positive/negative values.}
\end{figure}

\begin{figure}[tb]
 \includegraphics[width=0.48\textwidth]{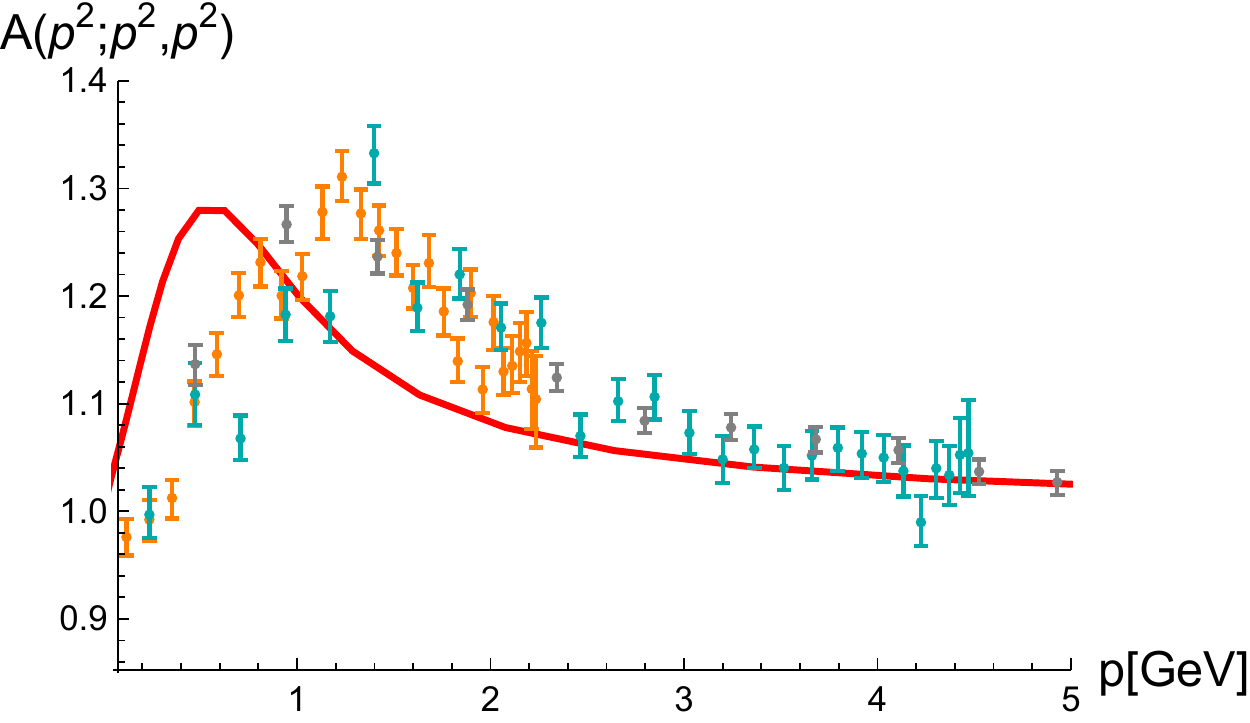}\\
\vskip2mm
 \includegraphics[width=0.48\textwidth]{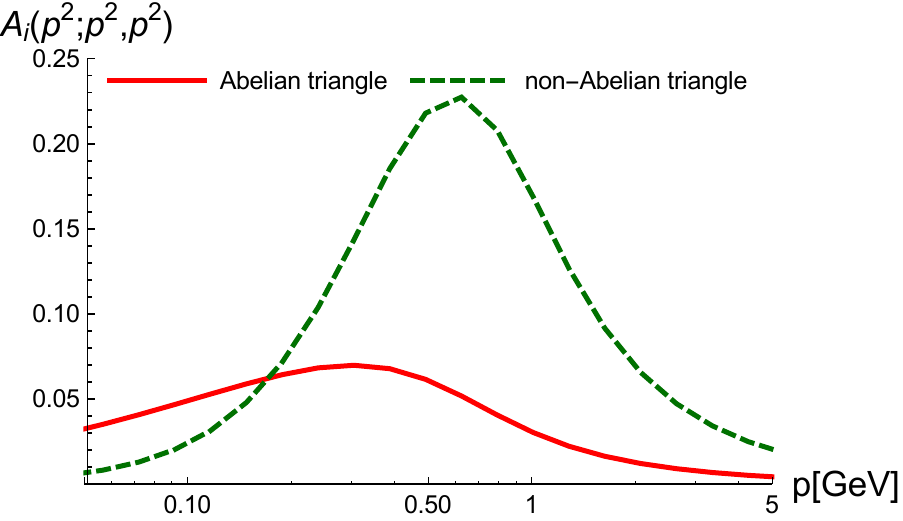}
 \caption{\label{fig:ghgWL_SD}Ghost-gluon vertex dressing from the full system in comparison with lattice results \cite{Maas:2016ip} (top) and the contributions from single diagrams (bottom). The shown lattice results correspond to lattice size $N=60$ and $\beta=3.18$, $5.61$ and $10.5$.} 
\end{figure}

\begin{figure}[tb]
 \includegraphics[width=0.48\textwidth]{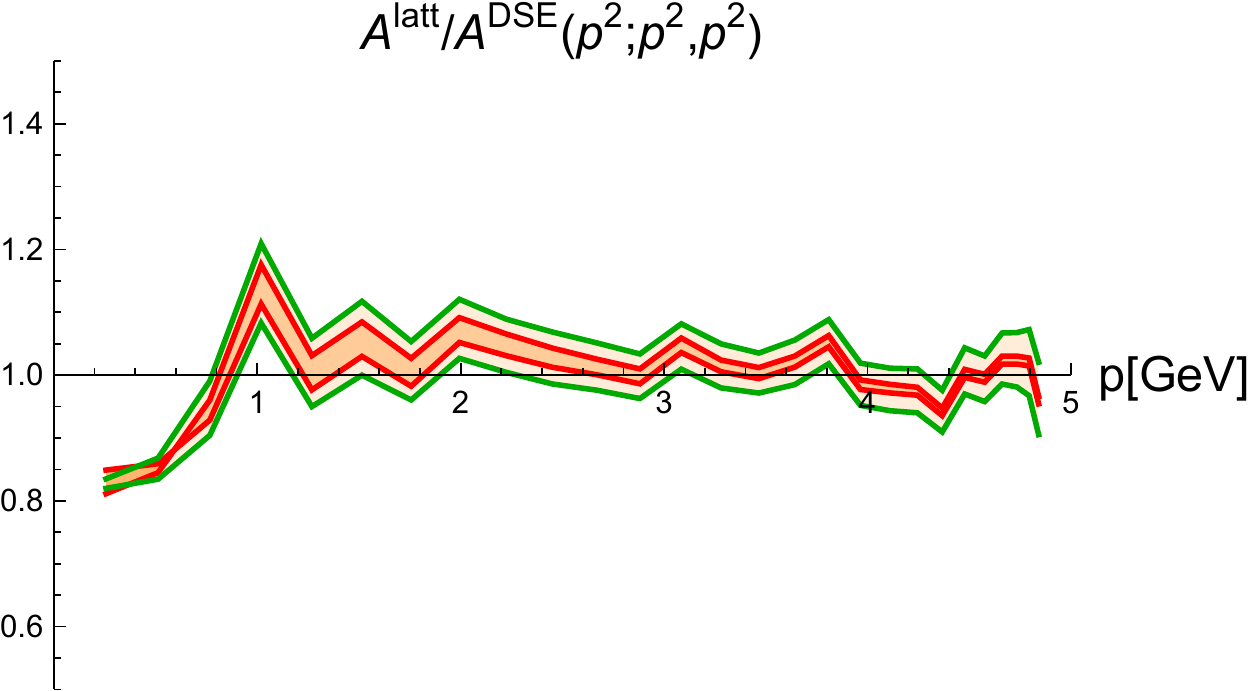}
 \caption{\label{fig:ghgRatio}Ratio of the ghost-gluon vertex dressing from the lattice ($N=60$, $\beta=10.5$) \cite{Maas:2016ip} over the DSE result. The red (inner) lines correspond to the results obtained from varying the scale, and the green (outer) lines represent the error from the lattice calculation.} 
\end{figure}

\begin{figure}[tb]
 \includegraphics[width=0.48\textwidth]{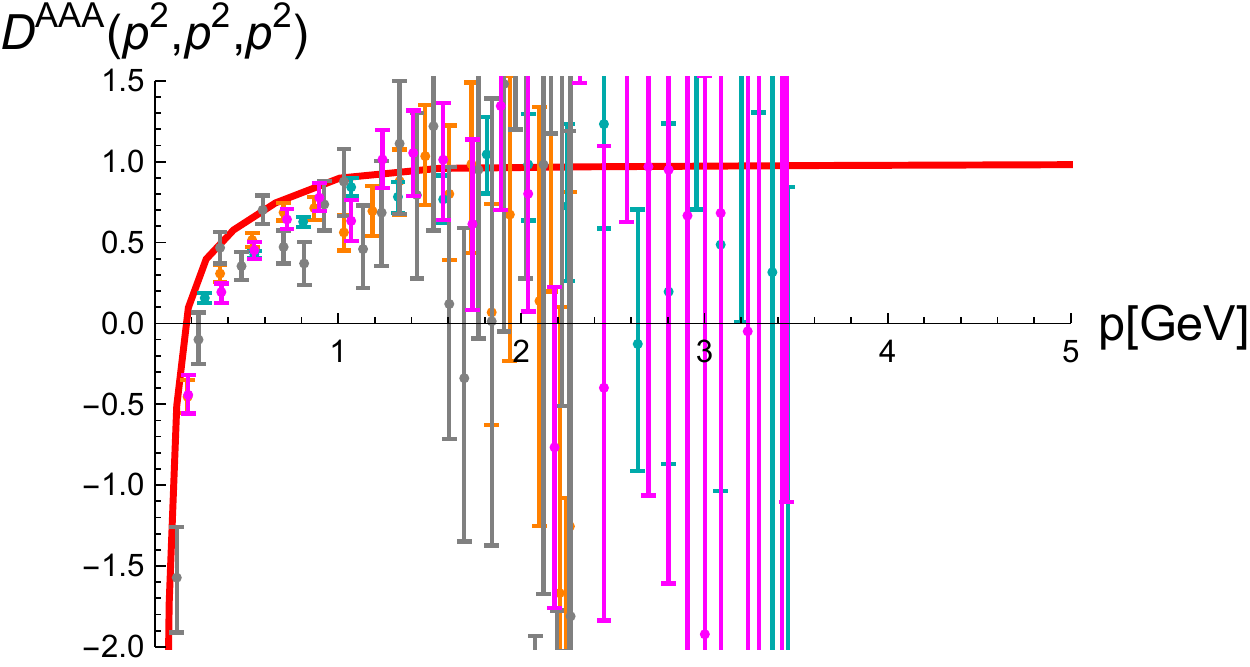}
  \vskip2mm
 \includegraphics[width=0.48\textwidth]{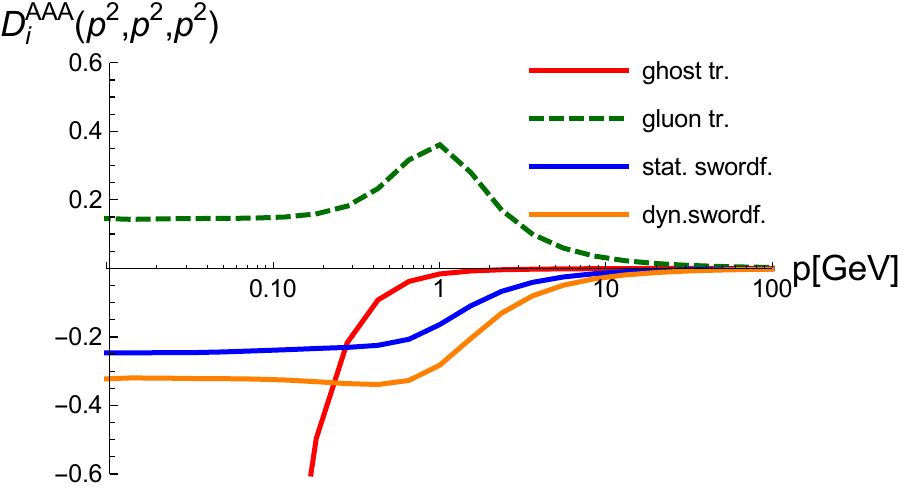}
 \caption{\label{fig:tgWL_SD}Three-gluon vertex dressing from the full system in comparison with lattice results \cite{Cucchieri:2008qm} (top) and the contributions from single diagrams (bottom). The shown lattice results correspond to lattice sizes $N=40$ and $60$ and $\beta=4.1$ and $6$.}
\end{figure}

\begin{figure}[tb]
 \includegraphics[width=0.48\textwidth]{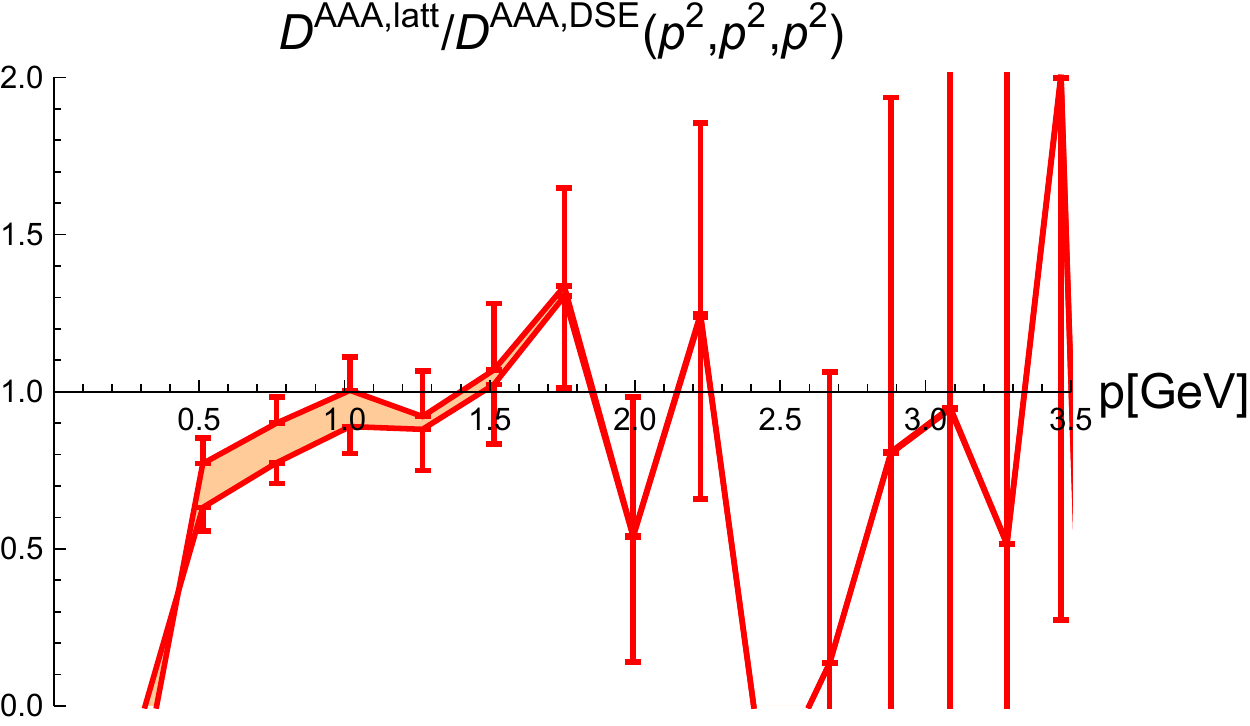}
 \caption{\label{fig:tgRatio}Ratio of the three-gluon vertex dressing from the lattice ($N=60$, $\beta=6$) \cite{Cucchieri:2008qm} over the DSE result. The connected points correspond to the results obtained from varying the scale. The error bars from the lattice are added on top of that.} 
\end{figure}

\begin{figure*}[tb]
 \includegraphics[width=0.48\textwidth]{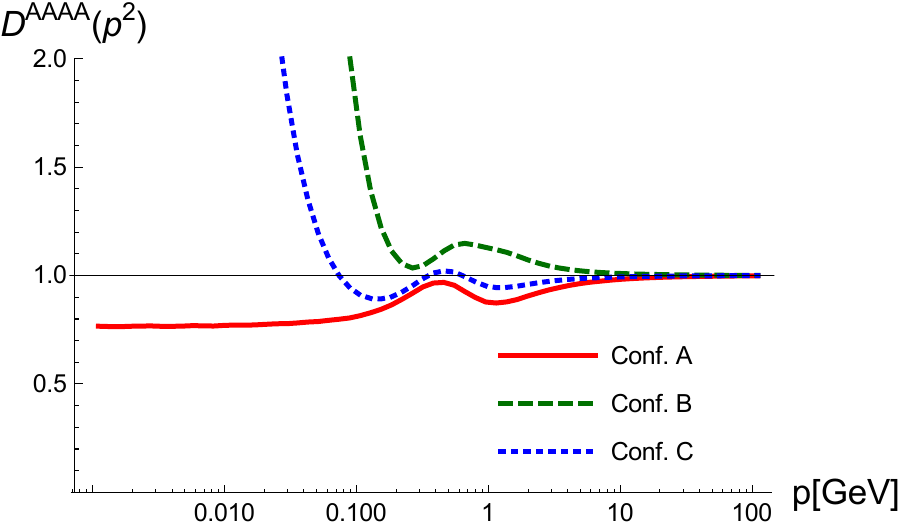}
 \hfill
 \includegraphics[width=0.48\textwidth]{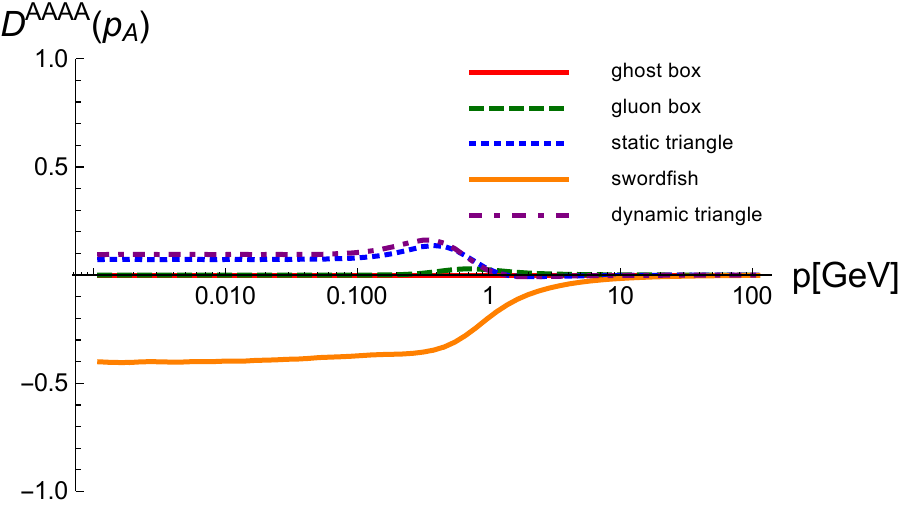}\\
 \vskip2mm
 \includegraphics[width=0.48\textwidth]{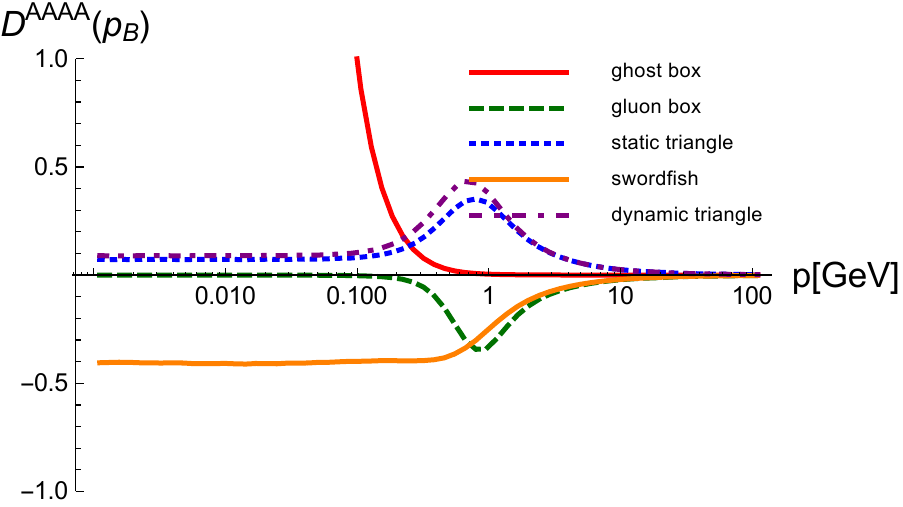}
 \hfill
 \includegraphics[width=0.48\textwidth]{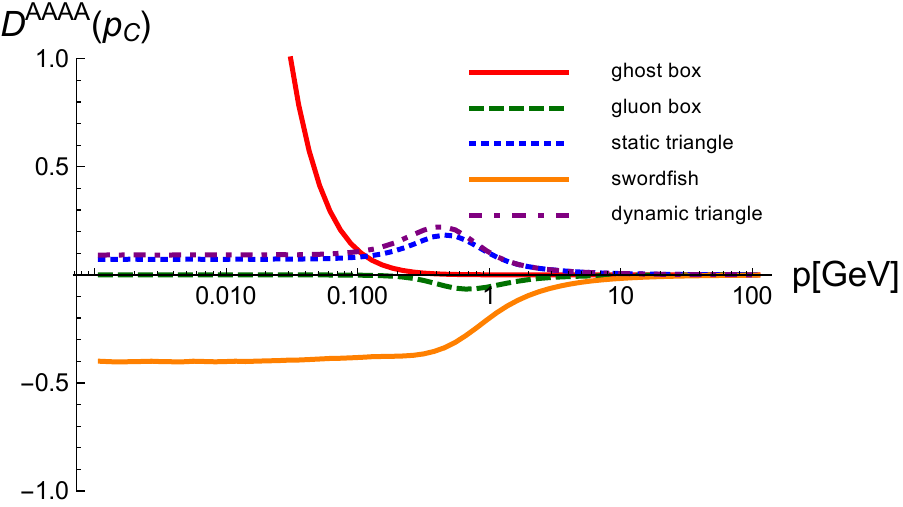}
 \caption{\label{fig:fg_V1}Tree-level dressing of the four-gluon vertex for different kinematic configurations (top left) and the individual contributions of single diagrams (top right and bottom).}
\end{figure*}

\begin{figure*}[tb]
 \includegraphics[width=0.48\textwidth]{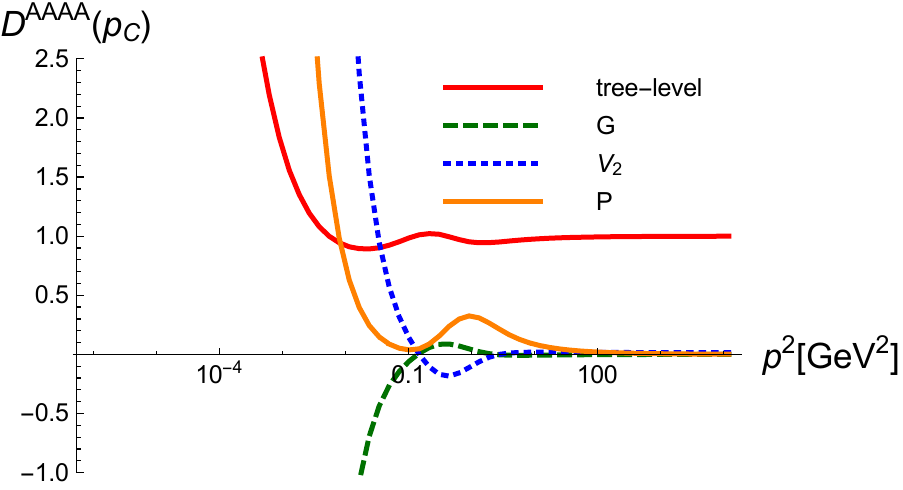}
 \hfill
 \includegraphics[width=0.48\textwidth]{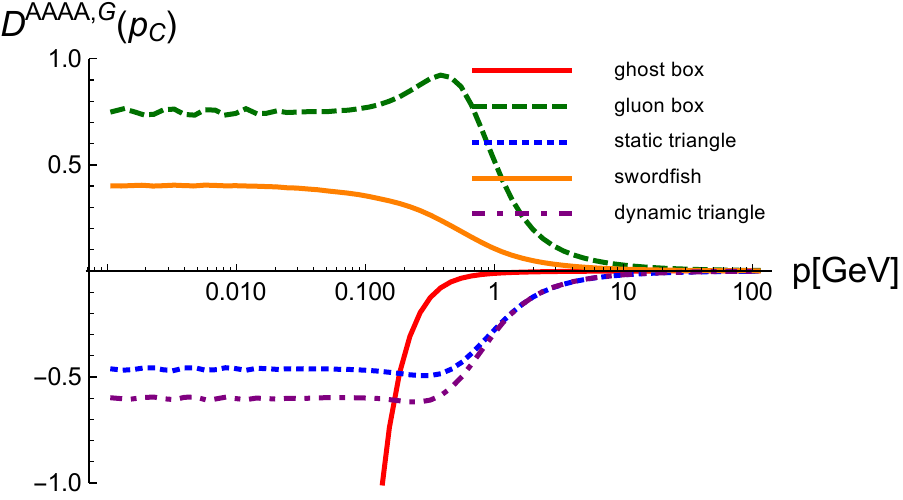}\\
 \vskip2mm
 \includegraphics[width=0.48\textwidth]{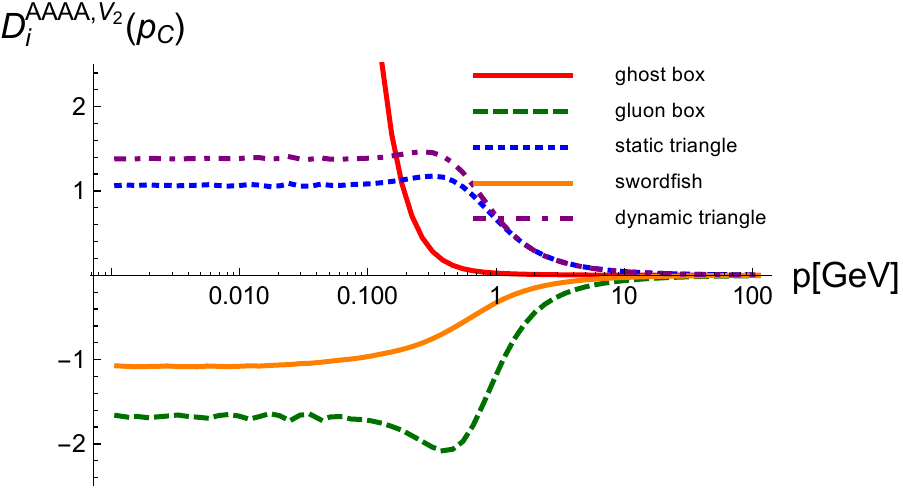}
 \hfill
 \includegraphics[width=0.48\textwidth]{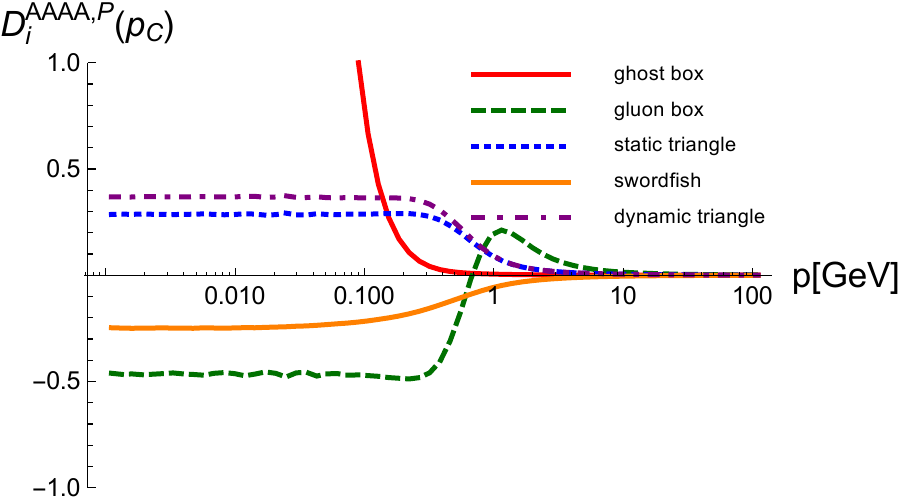}
 \caption{\label{fig:fg_other_proj}Various dressing functions of the four-gluon vertex (top left) and the individual contributions of single diagrams (top right and bottom).}
\end{figure*}

For the gluon dressing function an interesting question concerns the importance of single diagrams. To answer this, the contributions of each single diagram to the two-point function rescaled by $p^2$ are shown in \fref{fig:glDiagrams}. The spurious divergences were fitted for each diagram separately to obtain the finite contributions. As expected, the gluon loop is the dominant diagram in the UV. Also in the midmomentum regime it plays an important role. The ghost loop becomes important at low momenta. The tadpole can only contribute with a momentum independent constant. The two-loop diagrams also have a clear hierarchy with the squint diagram being more important than the sunset. Indeed the squint diagram yields the second largest contribution around $1\,\text{GeV}$. This explicitly confirms results from four dimensions \cite{Meyers:2014iwa}.

For the three-point functions results from the symmetric point configuration are shown. Other configurations are qualitatively similar.
The results for the ghost-gluon vertex are depicted in \fref{fig:ghgWL_SD}. The qualitative form is the same as that of the lattice results. However, the maxima of the two bumps do not coincide. The source of this deviation is currently not known. In \fref{fig:ghgWL_SD} also the contributions of the two diagrams are shown. Their importance depends on the specific kinematic configuration. The ratio of lattice and continuum results is depicted in \fref{fig:ghgRatio}. Again a band obtained from varying the scale is shown. In addition, the lattice error is added. It was found that for all configurations the deviation is always below $20\,\%$. In particular, above $1.5\,\text{GeV}$ the agreement is good.

The three-gluon vertex, shown in \fref{fig:tgWL_SD}, agrees well with the lattice data. The position of the zero crossing is in the deep IR. This is expected from lattice results. However, in four dimensions the position depends on the RG improvement employed for the three-gluon vertex DSE \cite{Eichmann:2014xya}. In three dimensions there is no reason to introduce such a term, and the fact that it is not required is consistent. The position of the zero crossing is close to the value of $134\,\text{Mev}$ determined in Ref. \cite{Aguilar:2013vaa}. Below the zero crossing the three-gluon vertex diverges linearly which confirms the findings of Refs.~\cite{Pelaez:2013cpa,Aguilar:2013vaa}. The ratio of lattice and continuum results is shown in \fref{fig:tgRatio}. Because of the large error bars at larger momenta, no band for the error is shown, but the error bars themselves are added on top of the results.

Finally, the results for the four-gluon vertex are discussed and are shown in Figs.~\ref{fig:fg_V1} and \ref{fig:fg_other_proj}. Three different kinematic configurations, indicated by $p_A$, $p_B$, and $p_C$, are shown; for details see Ref.~\cite{Cyrol:2014kca}. As can be seen in \fref{fig:fg_V1}, the swordfish and triangle diagrams yield the largest contributions. The ghost box is very small except for a linear IR divergence for configurations $B$ and $C$. The total tree-level dressing of the vertex shows that the deviations from a bare vertex are very small.

The complete four-gluon vertex is constructed from a basis of  $8$ color and $136$ Lorentz tensors, $41$ of which are transverse \cite{Eichmann:2015nra}. The calculation of the corresponding dressing functions is beyond the scope of this work. However, a few selected projections, see Ref.~\cite{Cyrol:2014kca} for details, are shown in \fref{fig:fg_other_proj}. The small wiggles observed for some diagrams are numeric artifacts. The qualitative picture is the same as for the tree-level dressing: The ghost box is small with a linear divergence in the IR and the gluon triangle and the swordfish diagrams yield large contributions. A notable difference is that the gluon box yields a larger contribution than for the tree-level dressing. Still, the sum of all diagrams is close to zero. This confirms the findings of Ref.~\cite{Cyrol:2014kca} that at least the investigated dressings beyond the tree-level dressing are small. Also, the IR divergence sets in at a very low scale. This explains the small influence of the four-gluon vertex on other correlation functions discussed in Sec.~\ref{sec:4g_effects}.

In summary, the deviation of the DSE results from the lattice results is below $20\,\%$ to a large extent. The largest deviation in terms of the ratio of the lattice over the continuum results was found for the ghost dressing function. Unfortunately, the possibility of different solutions does not allow one to draw any final conclusion about this deviation. However, a possible explanation for this might be the ghost-gluon vertex. It is clear from previous works \cite{Aguilar:2013xqa,Huber:2012kd} and the present work that the ghost-gluon vertex has a quantitative influence on the ghost propagator. Using a model for the vertex, it was tested that the ghost dressing function can indeed be changed, and this has also an effect on the strength of the gluon propagator. Finally, one should also note that the effects from different gauge fixing algorithms lie in the momentum regions where the deviation from lattice results is found to be largest: in the deep IR below $500\,\text{MeV}$ for the gluon propagator \cite{Bornyakov:2011fn,Maas:2013vd,Bornyakov:2013ysa} and for the ghost dressing function also at slightly higher momenta \cite{Maas:2013vd}.

\section{Summary and conclusions}
\label{sec:summary}

In this work the set of the Dyson-Schwinger equations for the five primitively divergent Green functions of three-dimensional Yang-Mills theory was investigated. The employed truncation prescription only drops diagrams with nonprimitively divergent Green functions leaving the propagator equations untouched and reducing the vertex equations to the UV leading one-loop diagrams. An important feature of this truncation is that it is self-contained and there is no freedom to model anything.

Three-dimensional Yang-Mills theory was chosen since it has several technical advantages over four dimensions which are related to the absence of renormalization. The most important one is that it was possible to extend the method for the subtraction of spurious divergences from Ref. \cite{Huber:2014tva} to include dynamic vertices and two-loop diagrams. How to do this is currently unknown for four dimensions, but this work provided some helpful insight. Another advantage is that there is no reason to introduce renormalization group improvement terms. In four dimensions such terms are necessary within typical truncations to obtain good UV properties, e.g., the correct anomalous dimensions of the propagators. However, they do have a quantitative influence as well. The observation that these terms do not have a large influence in three dimensions, where there is a priori no reason to introduce them, highlights once more that these terms should be better understood in four dimensions. In a nutshell one can say that the three-dimensional theory allows focusing on the investigation of pure truncation effects by avoiding a certain type of systematic uncertainties.

A main focus of this work was the analysis of the stability and reliability of truncating DSEs of Yang-Mills theory. Changing various points of the truncation like the employed equation for the ghost-gluon vertex showed that only small changes occur. As an alternative to DSEs, also the equations of motion from the 3PI effective action were used. At the employed level of truncations, the results turned out to be very similar. One reason for this is the four-gluon vertex, which is bare in the 3PI setup. The explicit DSE calculation showed that the deviations from the tree-level are indeed small.

It should be emphasized that for the gluonic vertices this is not the result of small contributions from single diagrams but comes from cancellations between the diagrams. In the three-gluon vertex the deviation from the tree-level starts around $1\,\text{GeV}$ and is mainly driven by the ghost triangle. The sum of the other diagrams, which are not small by themselves, is very small above $1\,\text{GeV}$. The situation for the four-gluon vertex is similar. The same was already observed in four dimensions \cite{Blum:2014gna,Eichmann:2014xya,Cyrol:2014kca}, but without the logarithmic running of the dressing function this behavior is even more pronounced and also independent from the question of RG improvement terms.
Note that for the ghost-gluon vertex no cancellations appear since -- within this truncation -- all contributions are positive.
  The presence of strong cancellations is of course interesting from the point of view of future truncations that might neglect sets of diagrams which sum up to a small contribution only. On the other hand, the question pops up about what the contributions of diagrams neglected in the present truncation are. If there is a similar cancellation mechanism at work, it could well be that adding only some of these diagrams could make the results worse again. The study of higher correlation functions could shed some light on this question.

The results obtained in this work improve in various ways our understanding of how to use functional equations in four dimensions. An important aspect is that it is expected that the hierarchy of the relevance of various Green functions is the same in three and four dimensions. Thus, in four dimensions the same truncation should yield quantitatively good results. The importance of various parts of the truncation was found to be as already known or conjectured in four dimensions. Especially the impact of two-loop diagrams in the gluon propagator DSE, in particular, of the squint diagram, is confirmed. For the three-gluon vertex, on the other hand, the results clearly indicate that a one-loop truncation is sufficient. Furthermore, the treatment of spurious divergences, which has a quantitative impact on the solutions, was shown to work also for dynamic vertices and two-loop diagrams. A generalization to four dimensions, which is complicated by the RG resummation, is a prerequisite for future studies. It should be noted that unquenching does not change the main conclusions of this work. For example, the gluon propagator DSE is unquenched by adding a single one-loop diagram which suppresses the gluon dressing function in the midmomentum regime; see, e.g., Refs.~\cite{Fischer:2003rp,Aguilar:2012rz}. For the three-point functions the effect is of the same magnitude or even lower \cite{Williams:2015cvx}.

In summary, the results of this work provide further evidence that functional equations are a reliable approach for the calculation of the elementary Green functions. These, in turn, provide access to physically relevant quantities like properties of different phases or bound states. Often such calculations rely on phenomenological models or use fits of lattice data. However, in cases where such fits are not available, for example at nonzero chemical potential, self-consistent calculations of the underlying quantities provide an alternative approach. Its applicability and feasibility in the vacuum were exemplified in this work.

\begin{acknowledgments}
I thank Gernot Eichmann, Axel Maas, Roman Rogalyov and Richard Williams for useful discussions. Furthermore, I am grateful for the lattice data provided by the authors of Refs.~\cite{Cucchieri:2008qm,Bornyakov:2013ysa,Maas:2014xma} and to Axel Maas for granting permission to show unpublished data and a critical reading of the manuscript.
Feynman diagrams were created with \textit{Jaxodraw} \cite{Binosi:2003yf}.
Results have been achieved using the Vienna Scientific Cluster (VSC) and the HPC clusters at the University of Graz.
Funding by the FWF (Austrian science fund) under Contract No. P 27380-N27 is gratefully acknowledged.
\end{acknowledgments}

\appendix

\section{Technical details}

Three-dimensional Yang-Mills theory is from the calculational point of view in many respects similar to the four-dimensional theory. Here I only refer to the relevant literature and note any differences.

The equations were derived with the \textsc{Mathematica} \cite{Wolfram:2004} application \textsc{DoFun} \cite{Alkofer:2008nt,Huber:2011qr}. For some algebraic manipulations of integral kernels \textsc{FORM} was used \cite{Vermaseren:2000nd}. In particular the integral kernels of the four-gluon vertex were optimized with the routines provided by \textsc{FORM} \cite{Kuipers:2013pba}. The framework of \textsc{CrasyDSE} was used for the numerical implementation \cite{Huber:2011xc}.
Since the DSEs are finite in three dimensions, the equations were not used in a subtracted form as often done in four dimensions.

In four dimensions the coupling is related to the scale via dimensional transmutation. In three dimensions the explicit mass dependence of the coupling sets the scale; see, e.g., Ref.~\cite{Teper:1998te}. Thus, any change in the coupling corresponds to a shift of the momentum scale, e.g., $G(p^2)|_{a\,g}=G(p^2/a^4)|_{g}$. To be explicit, all calculations were performed with $g=1$, and physical units were obtained by taking over the lattice scale via putting the peak of the gluon dressing at the same place; see Sec.~\ref{sec:results}.

For details on how to solve the vertices, see Refs.~\cite{Blum:2014gna,Cyrol:2014kca}. For the solution of the full system of equations a simple fixed-point iteration was employed. The gauge group $SU(2)$ was used throughout this work, as this is the gauge group also employed in most available lattice calculations in three dimensions.

The equations for the propagators can be found in many previous works. Specifically, the one-loop expressions can be found, e.g., in Refs.~\cite{Huber:2012kd,Huber:2014tva}. Only the integral measure $\int d^4q/(2\pi)^4$ must be replaced by
\begin{widetext}
\begin{align}
 \int \frac{d^3q}{(2\pi)^3} = \frac{1}{(2\pi)^3} \int dq\, q^2 \int_0^{2\pi} d\phi \int_0^\pi d\theta \sin \theta.
\end{align}
The two-loop terms in the gluon propagator DSE are given by
\begin{align}
 \Sigma_\text{squint}(p^2)&=\int \frac{d^3q_1}{(2\pi)^3}\int \frac{d^3q_2}{(2\pi)^3}K_\text{sq}(p,q_1,q_2) D^{A^3}(p^2, (p+q_1)^2, q_1^2)D^{A^3}(q_1^2, (q_1-q_2)^2, q_2^2)\nnnl
 & \quad \quad \times Z((p+q_1)^2)Z((q_1-q_2)^2)Z(q_1^2)Z(q_2^2),\\
 \Sigma_\text{sunset}(p^2)&=\int \frac{d^3q_1}{(2\pi)^3}\int \frac{d^3q_2}{(2\pi)^3}K_\text{su}(p,q_1,q_2) D^{A^4}(p^2, (q_1-q_2)^2, (p+q_1)^2, q_2^2))Z((p+q_1)^2)Z((q_1-q_2)^2)Z(q_2^2),
\end{align}
with the kernels (in $d$ dimensions)
\begin{align}
K&_\text{sq}(p,q_1,q_2)=3 \Big((4 q_1^2 q_2^2 (q_1^2-3 (q_1-q_2)^2-q_2^2) (p.q_1)^3+(p.q_1)^2 (q_1^2 q_2^2 ((p^2+3 q_1^2) (q_1^2-3 (q_1-q_2)^2-q_2^2)\nnnl
&+(p+q_1)^2 (-11 q_1^2+2 d\, q_1^2+(21-6 d) (q_1-q_2)^2+3 q_2^2-2 d\, q_2^2))+q_2^2 ((p^2-q_1^2) (q_1^2+(q_1-q_2)^2-q_2^2)\nnnl
&+(p+q_1)^2 (7 q_1^2-(q_1-q_2)^2+q_2^2)) q_1.q_2+(-p^2+q_1^2+(p+q_1)^2) (q_1^2+(q_1-q_2)^2-q_2^2) (q_1.q_2)^2\nnnl
&+4 q_1^2 p.q_2 (q_2^2 (-q_1^2-(q_1-q_2)^2+q_2^2)+2 (q_1-q_2)^2 q_1.q_2))+p.q_1 (-2 q_1^2 p.q_2 ((p^2+q_1^2+\nnnl
&(-3+d) (p+q_1)^2) q_2^2 (q_1^2+(q_1-q_2)^2-q_2^2)-((p^2+3 q_1^2) (q_1-q_2)^2+(p^2-q_1^2) (q_1^2-q_2^2)\nnnl
&+(p+q_1)^2 (3 q_1^2+(-9+2 d) (q_1-q_2)^2+q_2^2)) q_1.q_2+4 (p+q_1)^2 (q_1.q_2)^2)\nnnl
&+p^2 (q_1^2 q_2^2 (4 p^2 q_1^2-d\, p^2 q_1^2-8 q_1^4+d\, q_1^4+3 ((-2+d) p^2-(-6+d) q_1^2) (q_1-q_2)^2+d\, p^2 q_2^2+4 q_1^2 q_2^2-d\, q_1^2 q_2^2+\nnnl
&(p+q_1)^2 ((-4+d) q_1^2-3 (-2+d) (q_1-q_2)^2-d\, q_2^2))+q_2^2 (7 q_1^4+3 q_1^2 (p+q_1)^2+(p^2+3 q_1^2-(p+q_1)^2) (q_1-q_2)^2\nnnl
&-3 q_1^2 q_2^2+(p+q_1)^2 q_2^2-p^2 (3 q_1^2+q_2^2)) q_1.q_2+((-p^2-7 q_1^2+(p+q_1)^2) (q_1-q_2)^2\nnnl
&+(-p^2+q_1^2+(p+q_1)^2) (q_1^2-q_2^2)) (q_1.q_2)^2))+q_1^2 (q_1^2 (p^2-q_1^2+(p+q_1)^2) (-q_1^2+(q_1-q_2)^2+q_2^2) (p.q_2)^2\nnnl
&+p^2 (-p^2+q_1^2+(p+q_1)^2) p.q_2 (-(-2+d) q_2^2 (q_1^2+(q_1-q_2)^2-q_2^2)+2 (q_1^2+(-3+d) (q_1-q_2)^2+q_2^2) q_1.q_2\nnnl
&-4 (q_1.q_2)^2)+p^2 (q_1^2 q_2^2 (3 ((-1+d) p^2-(-5+d) q_1^2) (q_1-q_2)^2+q_1^2 ((-7+d) q_1^2-(-3+d) q_2^2)\nnnl
&+p^2 (3 q_1^2-d\, q_1^2+q_2^2+d\, q_2^2))+q_2^2 (d (p^2-q_1^2) (q_1^2+(q_1-q_2)^2-q_2^2)+4 q_1^2 (-p^2+2 q_1^2+(q_1-q_2)^2-q_2^2)) q_1.q_2\nnnl
&+(((3-2 d) p^2+(-11+2 d) q_1^2) (q_1-q_2)^2-(p^2-q_1^2) (3 q_1^2+q_2^2)) (q_1.q_2)^2+4 (p^2-q_1^2) (q_1.q_2)^3\nnnl
&+(p+q_1)^2 (-3 q_1^2 q_2^2 ((-5+d) q_1^2-3 (-3+d) (q_1-q_2)^2+q_2^2-d\, q_2^2)+3 q_2^2 ((-6+d) q_1^2+(-2+d) (q_1-q_2)^2\nnnl
&-(-2+d) q_2^2) q_1.q_2-3 (3 q_1^2+(-7+2 d) (q_1-q_2)^2+q_2^2) (q_1.q_2)^2+12 (q_1.q_2)^3))))\Big)/(4(d-1) p^4 q_1^4 (p+q_1)^4 (q_1-q_2)^4 q_2^4),\\
K&_\text{su}(p,q_1,q_2)=-\Big((3 q_2^2 (p.q_1)^3+(p.q_2)^4+(p.q_2)^3 (-q_1^2+2 q_1.q_2)+p^2 p.q_2 ((5-2 d) q_1^2 q_2^2\nnnl
&+(2 q_1^2+(-7+2 d) (q_1-q_2)^2-3 q_2^2+2 d\, q_2^2) q_1.q_2-4 (q_1.q_2)^2)+(p.q_2)^2 (q_1^4+((-2+d) p^2\nnnl
&+(-5+d) (p+q_1)^2) (q_1-q_2)^2-p^2 q_2^2+d\, p^2 q_2^2-2 (p+q_1)^2 q_2^2+d (p+q_1)^2 q_2^2-(p^2+q_1^2+3 (p+q_1)^2) q_1.q_2\nnnl
&+(q_1.q_2)^2)+(p.q_1)^2 ((-2+d) p^2 q_2^2+q_2^2 (3 q_1^2+(-5+d) (p+q_1)^2+(-5+d) (q_1-q_2)^2+q_2^2)+(p.q_2)^2\nnnl
&-4 q_2^2 q_1.q_2+(q_1.q_2)^2-p.q_2 (4 q_2^2+q_1.q_2))+p.q_1 (-2 (p.q_2)^3+(p.q_2)^2 (q_1^2+3 (q_1-q_2)^2+q_2^2-q_1.q_2)\nnnl
&+p.q_2 (-((-3+2 d) p^2+2 q_1^2+(-7+2 d) (p+q_1)^2) q_2^2+(p^2-2 q_1^2+3 (p+q_1)^2+3 (q_1-q_2)^2+q_2^2) q_1.q_2\nnnl
&+(q_1.q_2)^2)+p^2 (q_2^2 (-7 q_1^2+2 d\, q_1^2+2 (-5+d) (q_1-q_2)^2+2 q_2^2)+(q_2^2-2 d\, q_2^2) q_1.q_2+4 (q_1.q_2)^2))\nnnl
&+p^2 (q_2^2 ((-5+d) q_1^4+(-5+d) p^2 (q_1-q_2)^2+p^2 q_2^2)+(-2 p^2+(7-2 d) q_1^2) q_2^2 q_1.q_2+(p^2+3 q_1^2\nnnl
&+(-5+d) (q_1-q_2)^2-2 q_2^2+d\, q_2^2) (q_1.q_2)^2-3 (q_1.q_2)^3+(p+q_1)^2 (q_2^2 (3 (8-5 d+d^2) (q_1-q_2)^2\nnnl
&+(-5+d) q_2^2)-2 (-5+d) q_2^2 q_1.q_2+(-5+d) (q_1.q_2)^2)))\Big)/(2(d-1) p^4 (p+q_1)^4 (q_1-q_2)^4 q_2^4).
\end{align}
\end{widetext}

The kernels for the ghost-gluon vertex can be found in Ref.~\cite{Huber:2012kd}, where again the integral measure has to be changed accordingly. For the three- and four-gluon vertices the kernels become very long, and they are not shown explicitly.

\section{Spurious divergences}
\label{sec:spurDivs}

\begin{figure*}[tb]
 \begin{center}
  \includegraphics[width=0.48\textwidth]{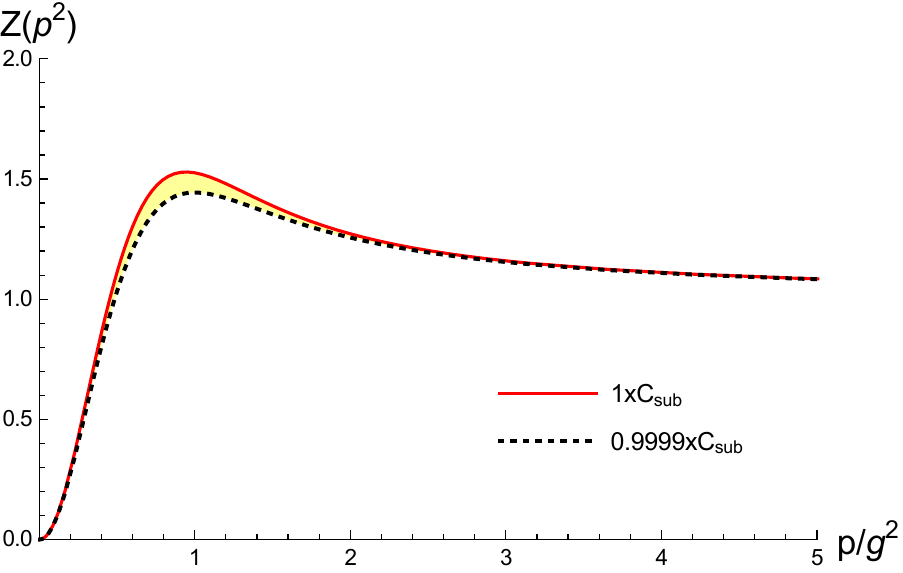}
  \hfill
  \includegraphics[width=0.48\textwidth]{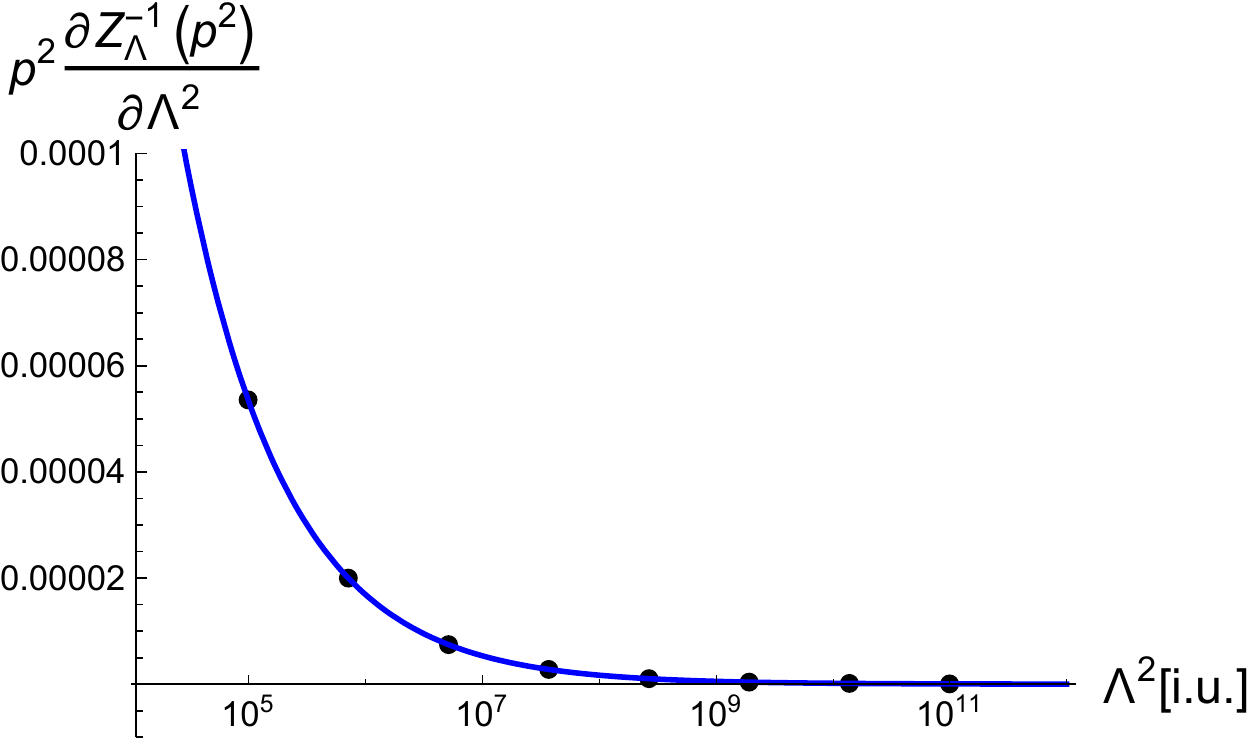}
  \caption{\label{fig:spurDivsSubcoeff}Left: Gluon dressing function calculated from the system of ghost and gluon propagators with a bare ghost-gluon vertex and a modeled three-gluon vertex with the correct and a rescaled value for $C_\text{sub}$. Right: Cutoff dependence of the right-hand side of the gluon propagator DSE. The dots correspond to calculated values, and the line corresponds to the fit function.}
 \end{center}
\end{figure*}

Spurious divergence in three dimensions have been explicitly treated in Ref. \cite{Maas:2004se} where they were subtracted via modifications of the integration kernels. To avoid modifying the IR behavior, the modification of the ghost loop included a compensating IR part. Since in that case only the scaling solution was investigated, simple power laws were sufficient. When employing this method with a decoupling solution, the problem arises that the leading IR behavior of the ghost is constant and it will give rise to spurious divergences. To circumvent the problem, a damping can be introduced, which, however, introduces an artificial scale. As was tested explicitly, the results depend on this scale.

Instead of this method, here the analytic calculation of Ref.~\cite{Huber:2014tva} is repeated for three dimensions. As in the four-dimensional case, it is not sufficient to calculate the UV behavior with bare dressings, but the one-loop corrections need to be taken into account,
\begin{align}
 G(p^2)&=1+\frac{g^2\,N_c}{16p},\\
 Z(p^2)&=1+\frac{11g^2\,N_c}{64p},
\end{align}
where $G(p^2)$ and $Z(p^2)$ are the dressing functions of the ghost and gluon propagators, respectively, and $N_c$ is the number of colors; see, e.g., Ref.~\cite{Maas:2004se}. While the trivial part leads to a linear divergence in the cutoff $\Lambda$ at order $g^2$, the $1/p$ part leads to a logarithmic dependence at order $g^4$. Using bare vertices in the UV, the divergent part of the gluon propagator DSE can then be calculated as
\begin{align}
 Z_\text{spur}(p^2,\Lambda)=\frac{C_\text{sub}}{p^2}=a\frac{g^2\,N_c\,\Lambda}{p^2} + b\frac{g^4\,N_c^2\,\log{\Lambda}}{p^2}.
\end{align}
The contributions of the single diagrams to the coefficients $a$ and $b$ are collected in Table~\ref{tab:spurDivsCoeffs}. Higher order terms are suppressed; e.g., the $g^6$ term is suppressed by $1/\Lambda$ as can be seen from dimensional arguments. To get rid of spurious divergences, $Z_\text{spur}(p^2,\Lambda)$ is subtracted in the gluon propagator DSE. It should be noted that this procedure works only if the employed model vertices approach their asymptotic form sufficiently fast. However, in contrast to four dimensions their leading correction does not need to be taken into account since they vanish polynomially in the UV. Note that the tadpole term can be included with this method. Its role for maintaining gauge invariance is also stressed, for example, by the fact that the subtraction coefficient at one-loop order is independent of the gauge fixing parameter in linear covariant gauges. 

\begin{table}[b]
 \begin{tabular}{|l||c|c||c|c|}
  \hline
   Ghost-loop & $a_\text{gh}$ & $\frac{1}{6\pi^2}$ &  $b_\text{gh}$ & $\frac{1}{48\pi^2}$\\
  \hline
  Gluon-loop & $a_\text{gl}$ & $-\frac{2}{3\pi^2}$ & $b_\text{gl}$ & $-\frac{11}{48\pi^2}$\\
  \hline
  Tadpole & $a_\text{tad}$ & $\frac{2}{3\pi^2}$ & $b_\text{tad}$ & $\frac{11}{96\pi^2}$\\
  \hline
 \end{tabular}
 \caption{\label{tab:spurDivsCoeffs}Contributions from the one-loop diagrams to the coefficients $a$ and $b$.}
\end{table}

As in four dimensions, analytic calculations no longer work when numerically calculated vertices are used. However, the coefficients $a$ and $b$ can be fitted. To get rid of the parts that do not depend on the cutoff, the derivative of the self-energy with respect to the cutoff is calculated. This procedure turns out to be stable not only for use with dynamic vertices, but also for the two-loop diagrams of the gluon DSE. It was used throughout this work. To illustrate the importance of the subtraction term $C_\text{sub}$, \fref{fig:spurDivsSubcoeff} shows the effect when it is lowered to $0.9999$ of the correct value. It is noteworthy that the value cannot be raised since then the gluon propagator DSE no longer converges and the bump gets higher in each iteration. Thus, the fitted value seems to correspond to a maximal value.

\bibliographystyle{utphys_mod}
\bibliography{literature_YM_3d}

\end{document}